\newif\ifextended
\extendedtrue
\documentclass[acmsmall,screen,nonacm]{acmart}

\usepackage{tikz}
\usepackage{tikz-cd}
\usepackage{mathpartir}
\usepackage{cleveref}

\newtheorem*{theorem*}{Theorem}
\theoremstyle{remark}

\theoremstyle{example}

\usetikzlibrary{decorations.pathmorphing}
\usetikzlibrary{arrows.meta}

\ifdefined\ifextended\else\expandafter\newif\csname ifextended\endcsname\fi
\newcommand{\appref}[2]{Appendix~\ifextended\ref{#1}\else#2 of the extended version\fi}
\providecommand{\xrp}{} %

\newcommand{\ret}{\mathsf{ret}}

\newcommand{\thunk}{\mathsf{thunk}}
\newcommand{\force}{\mathsf{force}}

\newcommand{\unit}{\mathsf{1}}
\newcommand{\zero}{\mathsf{0}}

\newcommand{\triv}{\langle\rangle}
\newcommand{\unitv}{()}

\newcommand{\with}{\;\&\;}

\newcommand{\op}{\mathsf{op}}
\newcommand{\id}{\mathsf{id}}
\newcommand{\ops}{\mathsf{Ops}}
\newcommand{\arity}{\mathsf{ar}}

\newcommand{\case}{\mathsf{case}}
\newcommand{\var}{\mathsf{var}}
\newcommand{\ctx}{\mathsf{ctx}}
\newcommand{\lhs}{\mathsf{lhs}}
\newcommand{\rhs}{\mathsf{rhs}}
\newcommand{\eq}{\mathsf{eq}}
\newcommand{\eqns}{\mathsf{Eqns}}

\newcommand{\Term}[2]{\mathsf{Term}_{#1}(#2)}
\newcommand{\alg}[1]{\mathsf{Alg}_{#1}}

\newcommand{\Alg}[1]{\mathsf{ALG}_{#1}}

\newcommand{\Model}[1]{\mathsf{MOD}_{#1}}
\newcommand{\den}[1]{\llbracket #1 \rrbracket}

\newcommand{\Set}{\mathsf{Set}}
\newcommand{\Sh}{\mathsf{Sh}}            %
\newcommand{\cont}[2]{#1 \lhd #2}        %
\newcommand{\pr}{\mathsf{pr}}            %
\newcommand{\dT}{\partial_T}             %
\newcommand{\config}[2]{\langle #1, #2\rangle} %
\newcommand{\sigmab}{{\sigma_\bullet}}      %
\newcommand{\prb}{{\pr_\bullet}}            %
\newcommand{\pbul}{p_\bullet}             %
\newcommand{\sbul}{s_\bullet}             %
\newcommand{\inl}{\mathsf{inl}}
\newcommand{\inr}{\mathsf{inr}}

\newcommand{\prop}{\mathsf{Prop}}

\newcommand{\mtree}{\mapsto_\mathsf{tree}}
\newcommand{\mS}{\mapsto_\mathsf{S}}
\newcommand{\thy}[1]{\mathcal{T}_{\textrm{#1}}}   %
\newcommand{\sig}[1]{\Sigma_{\textrm{#1}}}        %
\newcommand{\mT}{\mapsto_\mathsf{T}}

\newcommand{\probchoice}[3]{#1\mathop{+_{#3}}#2}

\DeclareRobustCommand{\grayhl}[1]{%
  \colorbox{gray!20}{\strut #1}%
}

\begin{document}
\title{An Algebraic Perspective on Operational Semantics}
\ifextended\titlenote{Extended version, with appendices, of the paper to appear at POPL 2027.}\fi

\author{Eric Bond}
\affiliation{
  \department{Electrical Engineering and Computer Science}
  \institution{University of Michigan}
  \country{USA}
}
\email{bonderic@umich.edu}
\orcid{0009-0008-1575-9606}

\author{Max S. New}
\affiliation{
  \department{Electrical Engineering and Computer Science}
  \institution{University of Michigan}
  \country{USA}
}
\email{maxsnew@umich.edu}
\orcid{0000-0001-8141-195X}
\begin{abstract}
Languages with effects are typically given two independently 
specified semantic accounts: a denotational semantics based on (free) models,
and an operational semantics defined by effect-specific 
reduction rules. This separation leads to two independently constructed 
semantic structures whose relationship must then be established a posteriori.
We present a unified construction in which both denotational interpretation 
and operational reduction are constructed from an algebraic theory specifying 
the effects.

We work with call-by-push-value extended by the operations of an
algebraic theory. We prove equational canonicity and give a tree-based
operational semantics that reduces every closed computation to a tree of
operations. When the free model monad is a monadic container, we derive
from it a configuration-based small-step semantics. Configurations are
elements of the free model. Every step preserves the denotation. Under a
condition we call affinity, every reduction terminates and at ground types
computes the denotation. We instantiate the framework with state, errors,
writer, and weighted and pre-convex probabilistic choice. For errors,
writer and weighted choice we recover the familiar rules. For state we get
a machine that runs the program from every initial state at once.
\end{abstract}
\maketitle

\section{Introduction}
Operational semantics and denotational semantics offer complementary views of 
effectful computation. Denotational semantics provides a mathematical 
account of effects, using monads and their corresponding algebraic theories, 
while operational semantics explains execution through computable reduction 
rules. Although these two perspectives are intended to describe the same 
computations, they are typically developed independently.

Moggi systematized the denotational approach to effects by modeling
effectful programs in the Kleisli category of a monad \cite{Monads-Moggi}.
Plotkin and Power refined this approach with ``algebraic effects'': by
viewing the monads as the free models of an algebraic theory, we get a
direct way to connect the effects with the primitive ``effectful
operations'' and axioms that are given in an effectful calculus
\cite{Notions-Computation-Monads-Plotkin-Power}. The algebraic perspective has the benefit that effects and
their laws can be studied individually and combined together
compositionally using operations such as coproduct and tensor of
theories \cite{HYLAND200670}. This provides a rich
toolkit for specification and reasoning about effects.

On the other hand, the analysis of operational semantics of these same
effects is less unified. Operational semantics of algebraic effects
originally computed a tree of operations and left them uninterpreted
\cite{Plotkin-Power-Adequacy,Generic-Op-Metatheory-Alg-Effect-Johann}.
Categorical approaches using coalgebra \cite{GSOS-Turi-Plotkin} now
extend to higher-order languages \cite{GSOS-Binders}. They give general
results about operational properties like bisimulation, but each
language still gives the rules for its effects by hand, and state needs a
new approach \cite{GSOS-State}.

Our goal in this paper is to bring the benefits of the systematic
perspective of algebraic effects to the operational semantics of
effects. A single reduction rule handles every operation, and for a
particular algebraic theory it yields the reduction rules of that
theory's effects. That is, we
reproduce the operational semantics used in practice by \emph{deriving}
them from an algebraic theory. 
The power of the systematic approach is that then soundness, adequacy and
logical relations arguments for these operational systems can be
provided once and for all, with a completely uniform development.

The main concept we need to formalize is the runtime
state of the operational semantics, traditionally called a
\emph{configuration} \cite{Plotkin-SOS}. For a language that logs messages, a configuration pairs the running
program with the log written so far. We
take configurations to be elements of the free model of the effect
theory. The unit of the free model monad gives the initial configuration
of a term, and the multiplication of the monad merges the result of interpreting an
operation, reached by one of the terms, back into the configuration.
Earlier operational semantics also place programs inside a monad
\cite{Gavazzo-Faggian-MonadicRewriting,GTV-MonadicIntersection,DGZ-MonadicTypeEffect}.
The idea is to define reduction as a relation
$R \subseteq \mathsf{Term} \times T(\mathsf{Term})$ and to lift it to
$T(\mathsf{Term})$ by Kleisli extension. So in all these works, a step
reduces every term inside the monad at once. We reduce one term at a
time, as the operational semantics used in practice do, and prove that
every step preserves the denotation of the configuration.

Our construction requires the free model to be a
monadic container
\cite{Uustalu-ContainerCombinatorics,Purdy-Damato-MonadicContainers}, a
monad whose underlying functor is a container, also known as a polynomial functor
\cite{Containers-ABBOTT20053,polynomial-GAMBINO_KOCK_2013}. Each
configuration is then a shape, the state of the machine, together with
the terms at its positions. The effect rules are computed from this
data. Configuration reduction need not terminate in general. We introduce
a sufficient condition, which we call \emph{affinity}. A container is
affine when a step never copies the terms around the one it reduces.
State, errors, writer, and weighted and pre-convex probabilistic
choice all give affine monadic containers. Finite nondeterminism and
finitely supported probability distributions do not give monadic
containers at all, since their free models are not container functors
(\Cref{sec:non-container}).
How an operation changes the configuration, its new state and where each
term goes, is computed from the free model alone, independently of the syntax
of the language.

The relation then between the denotational and operational semantics
within our framework becomes clear. In a denotational semantics, the
effects of a program are evaluated ``bottom-up'': in an effectful
operation $\mathsf{op}(M,\ldots)$ we first take the denotations
$\den{M},\ldots$ and then combine them together with the semantic
interpretation of the operation. Contrariwise, in an operational
semantics, the effects of a program are evaluated ``top-down''. The
configuration records the effects performed so far, and when one of its
subterms reaches an operation $\mathsf{op}(M,\ldots)$, the operation can be
interpreted right away, before any of its operands has run. The operands
$M,\ldots$ become new subterms of the configuration, to be run later.
For example, with writer effects a configuration $\langle m, N\rangle$
pairs the log $m$ with a term $N$. The step
$\langle m , \mathsf{tell}_n(N)\rangle \mapsto \langle m \otimes n , N \rangle$
records $n$ before $N$ runs. Take $M = \mathsf{tell}_a(\mathsf{tell}_b(\ret\,x))$
and the empty log $e$. The denotation adds $b$ to the log first but the
machine adds $a$ first.
\[
\begin{array}{@{}r@{\;}l@{\qquad\qquad}r@{\;}l@{}}
\den{M} & = \mathsf{tell}_a(\mathsf{tell}_b(\den{\ret\,x})) &
  & \langle e, \mathsf{tell}_a(\mathsf{tell}_b(\ret\,x)) \rangle \\
  & = \mathsf{tell}_a(\mathsf{tell}_b((e, x))) &
\mapsto & \langle e \otimes a, \mathsf{tell}_b(\ret\,x) \rangle \\
  & = \mathsf{tell}_a((b \otimes e, x)) &
\mapsto & \langle (e \otimes a) \otimes b, \ret\,x \rangle \\
  & = (a \otimes (b \otimes e), x) & &
\end{array}
\]
In this example we can easily see that the two results agree by the monoid laws. We generalize this work to more general monads.
Since the behavior of both approaches is derived from the same notion of a
free model, a short calculation with the monad laws and the fact
that stacks are interpreted as homomorphisms shows that every step of
the operational semantics preserves the denotation. Under affinity, at
ground types, every run of the machine ends in the denotation of the
program, which is computational adequacy.

We instantiate this approach concretely by formalizing a
Call-by-push-value (CBPV) calculus parameterized by an algebraic
theory. We use CBPV as a simple metalanguage into which call-by-value
and call-by-name calculi can be easily embedded
\cite{Levy-adjunction-model}. The calculus can be straightforwardly
given a denotational semantics following Levy's original algebra
models. We then give two effect-generic logical relations models
establishing canonicity of the equational theory (every closed 
computation of type $FA$ is provably equal to a tree of operations with 
returned values at its leaves) and termination of a
tree-based operational semantics that leaves algebraic effects
uninterpreted, as in effect trees and interaction trees
\cite{Generic-Op-Metatheory-Alg-Effect-Johann,Guarded-Interaction-Trees,Adequacy-Revisited-Kavvos}. We
provide a categorical analysis of how the equational canonicity proof
is reached. We then show that, given a monadic container that is the 
free model of the theory, we can
derive a configuration-based operational semantics, and we use the
affinity side condition to transport the termination proof from the
tree semantics to the configuration semantics.

\paragraph{Contributions.}
\begin{itemize}
\item Following standard techniques, we prove equational and operational
canonicity for CBPV with a finitary algebraic theory, using effect-generic
logical relations
(\Cref{sec:denotational-semantics-equational-canonicity,sec:operational-canonicity-tree}).
We write out in detail material that is scattered across the CBPV
literature and folklore.
\item For any algebraic theory whose free model is a monadic container, we
derive an operational semantics whose configurations are elements of the
free model (\Cref{sec:config-reduction}). A single rule, computed from the
monad structure of the container, interprets every operation, acting on one
subterm of a configuration at a time.
\item We prove that each step of this configuration semantics preserves the
interpretation of a configuration in the free model, given by the standard
denotational semantics of CBPV. When the container is
affine, we also prove strong normalization and computational adequacy at
ground types (\Cref{thm:soundness-mT,thm:sn-T,thm:adequacy}).
\item We instantiate the construction for errors, writer, finite state, and
weighted and pre-convex probabilistic choice (\Cref{sec:examples}). For
errors, writer and weighted choice we recover the usual rules, and for
finite state we obtain a machine that runs the program from every initial
state at once.
\end{itemize}

\section{Algebraic Effects in CBPV}\label{sec:Syntax}
In this section we describe a simply typed call-by-push-value (CBPV) calculus 
extended with primitive operations from a finitary algebraic theory of effects. 
The syntax distinguishes values, computations, and stacks in the standard 
CBPV style, and includes algebraic operations as term formers subject to the 
equations of the theory.
\subsection{Algebraic Theories and Algebraic Effects}
\label{sec:algebraic-theories}

Algebraic effects~\cite{Alg-Effects-Plotkin-Power} provide a way of
describing computational effects such as state, exceptions,
nondeterminism, and I/O using algebraic structure. The key idea is
that effectful programs can be viewed as computations generated from
primitive operations (e.g.\ reading a value, updating state, or
raising an exception), together with equations governing their
interaction. Such computations are reduced to trees of operations,
which are later interpreted in semantic models.  This intuition is
formalized by the notion of an algebraic
theory~\cite{Lawvere-Theories}.  Algebraic effects refine the monadic
approach to computational effects \cite{Monads-Moggi}. The operations
and equations of the theory correspond to primitive constructs and
axioms of the calculus. The monad of the theory sends a set $X$ to
the free model of the theory on $X$, and the category of models of
the theory is isomorphic to the category of Eilenberg--Moore algebras of the monad.

We use a standard definition of finitary algebraic theory $\mathcal{T} = (\Sigma, \eqns)$,
consisting of a signature $\Sigma$ and a set of equations $\eqns$.
A \emph{signature} $\Sigma = (\ops, \arity)$ specifies a set of operations $\ops$ and for each operation $\op : \ops$, 
an arity $\arity(\op) : \mathbb{N}$. Given a signature $\Sigma$, 
the set of \emph{terms} generated by the signature with variables of type $X$ 
is inductively defined by: 
  \begin{mathpar}
    \inferrule 
      { x : X} 
      { \var\; x : \Term \Sigma X}
    \and
    \inferrule
      { \op : \ops \and \forall i.\;t_i : \Term \Sigma X}
      { \op(t_1,t_2,\ldots,t_{\arity(\op)}) : \Term \Sigma X}
  \end{mathpar}
 which we denote as $\Term \Sigma X$. 
 An \emph{equation} consists of a context size $\ctx : \mathbb{N}$ with two terms
 $\lhs, \rhs : \Term \Sigma \ctx$ where $\ctx$ is interpreted as a finite set of  cardinality $\ctx$.
 
 \begin{example}
  [Theory of Writer]
  The theory of writer \cite{HYLAND200670} models computations that accumulate output, such as
  logging, tracing, or incrementally constructing an output buffer.
  The type of output is represented by a monoid $(M,\otimes,e)$, where
  $e$ denotes the empty output and $\otimes$ combines the outputs produced
  by successive computations.
  For a given monoid, we define the theory 
  $\thy{writer}$ to consist of a signature $\sig{writer}$
  with one unary operation $\mathsf{tell}_m(t)$ for every $m : M$. 
  The equations of this theory are:
  \begin{mathpar}
  \inferrule
    {}
    {\mathsf{tell}_e(t) = t}
  \and  
  \inferrule 
    {}
    {\mathsf{tell}_m(\mathsf{tell}_n(t)) = \mathsf{tell}_{m \otimes n}(t)}
\end{mathpar}

\end{example}
\begin{example}
  [Theory of Errors]
  The theory of errors models computations that may abort~\cite[Ex.~1]{Notions-Computation-Monads-Plotkin-Power}. We define the
  theory $\thy{err}$ to consist of a signature $\sig{err}$
  with a single nullary operation $\mathsf{err}$, and no equations.
\end{example}
\begin{example}
  [Theory of Finite State]
  The theory $\thy{state}$ of state for a finite set of states $S$ consists of a 
  signature $\Sigma_S$ with unary operations $\mathsf{set}_s(t)$ for all $s : S$  
  and a single operation $\mathsf{get}(\overrightarrow{t})$ of arity $|S|$.
  The equations of $\thy{state}$ \cite[\S 3]{Notions-Computation-Monads-Plotkin-Power} consist of the usual 
\textit{lens laws}, for all $s, s' : S$:
\begin{mathpar}
  \inferrule
    {}
    {\mathsf{get}(\overrightarrow{\mathsf{set}_s(t)}) = t}
  \and
  \inferrule
    {}
    {\mathsf{set}_s(\mathsf{get}(\overrightarrow{t})) = \mathsf{set}_s(t_s)}
  \and  
  \inferrule 
    {}
    {\mathsf{set}_s(\mathsf{set}_{s'}(t)) = \mathsf{set}_{s'}(t)}
  \and
  \inferrule
    {}
    {\mathsf{get}(\overrightarrow{\mathsf{get}(\overrightarrow{t_s})}) = \mathsf{get}(\overrightarrow{t_{ss}})}
\end{mathpar}
\end{example}
\begin{example}
  [Theory of Weighted Monoids]
  The theory of weighted monoids models computations that may produce
  multiple weighted outcomes. Unlike ordinary nondeterminism, the order of
  outcomes is significant and duplicate outcomes are preserved. Let
  $(R,\cdot,1_R)$ be a multiplicative monoid of weights.
  We define the theory 
  $\thy{wmon}$ to consist of a signature $\sig{wmon}$
  with a nullary operation $e$, a binary operation $\otimes$, 
  and one unary operation $\mathsf{act}_r$ for each $r : R$. 
  The equations of this theory are given below.
  \begin{mathpar}
    \inferrule
      {}
      {t \otimes (t' \otimes t'') = (t \otimes t') \otimes t''}
    \and 
    \inferrule 
      {}
      {e \otimes t = t}
    \and  
    \inferrule 
      {}
      {t \otimes e = t}
    \and  
    \inferrule 
      {}
      {\mathsf{act}_{1_R}(t) = t}
    \and  
    \inferrule  
      {}
      {\mathsf{act}_r(\mathsf{act}_{r'}(t)) = \mathsf{act}_{r \cdot r'}(t)}
    \and  
    \inferrule 
      {}
      {\mathsf{act}_r(t \otimes t') = \mathsf{act}_r(t) \otimes \mathsf{act}_r(t')}
    \and  
    \inferrule 
      {}
      {\mathsf{act}_r(e) = e}
  \end{mathpar}
  Equivalently, a weighted monoid is a monoid $(M, \otimes, e)$ with an
  action of the multiplicative monoid $(R, \cdot, 1_R)$ by monoid
  homomorphisms. That is, $\mathsf{act}_1$ is the identity,
  $\mathsf{act}_r \circ \mathsf{act}_{r'} = \mathsf{act}_{r \cdot r'}$,
  and each $\mathsf{act}_r$ preserves $\otimes$ and $e$. This makes the
  theory of weighted monoids the tensor of the theory of monoids with
  the theory of writer for $(R, \cdot, 1_R)$ \cite{HYLAND200670}. Its
  free model is the writer monad transformer for $(R, \cdot, 1_R)$ applied
  to the list monad, $\mathsf{List}(R \times X)$
  \cite[Thm.~12]{HYLAND200670}.
\end{example}

\begin{example}
  [Theory of $R$-Semimodules]
  The theory of $R$-semimodules \cite[Ch.~14]{Golan-Semirings} models computations that can be combined into
  formal linear combinations. Here the semiring $R$ provides coefficients
  describing quantitative information such as multiplicities or costs.
  By choosing different semirings, one recovers familiar computational
  structures, including finite sets (Boolean semiring) and finite
  multisets (natural numbers). Formally, let $(R,+,\cdot,0,1)$ be a semiring. We define the theory
  $\thy{$R$-Semi}$ to consist of a signature
  $\sig{$R$-Semi}$ with:
  \begin{itemize}
    \item a constant $0$,
    \item a binary operation $+$,
    \item for each $r \in R$, a unary operation $\mathsf{scale}_r(-)$.
  \end{itemize}

  Intuitively, $+$ and $\mathsf{scale}_r$ together form linear
  combinations of computations. $+$ adds two computations, $0$
  represents the empty combination, and $\mathsf{scale}_r(t)$ scales the
  computation $t$ by the coefficient $r$.
  The equations of $\thy{$R$-Semi}$ are:
  \begin{mathpar}
    \inferrule
      {}
      {t + (t' + t'') = (t + t') + t''}
    \and
    \inferrule
      {}
      {t + t' = t' + t}
    \and
    \inferrule
      {}
      {t + 0 = t}
    \and
    \inferrule
      {}
      {\mathsf{scale}_1(t) = t}
    \and
    \inferrule
      {}
      {\mathsf{scale}_r(\mathsf{scale}_{r'}(t))
        = \mathsf{scale}_{r \cdot r'}(t)}
    \and
    \inferrule
      {}
      {\mathsf{scale}_r(t + t')
        = \mathsf{scale}_r(t) + \mathsf{scale}_r(t')}
    \and
    \inferrule
      {}
      {\mathsf{scale}_{r+r'}(t)
        = \mathsf{scale}_r(t) + \mathsf{scale}_{r'}(t)}
    \and
    \inferrule
      {}
      {\mathsf{scale}_0(t) = 0}
  \end{mathpar}
\end{example}

\begin{example}
  [Probability, Theory of Convex Spaces]
  The theory of convex spaces \cite{Stone-Barycentric,Fritz-ConvexSpaces} axiomatizes a notion of biased coin flip.
  For every probability $p \in [0,1]$ we have a binary operation
  \[ \probchoice{t_1}{t_2}{p} \]
  that we interpret as flipping a coin with probability $p$ and executing $t_1$ if true and $t_2$ if false.
  The equations are natural probabilistic laws:
  \begin{mathpar}
    \inferrule
      {}
      {\probchoice{t_1}{t_2}{1} = t_1}\and
    \inferrule
      {}
      {\probchoice{t_1}{t_2}{0} = t_2}\and
    \inferrule{}{\probchoice{t}{t}{p} = t}
    \and
    \inferrule{}{\probchoice{t_1}{t_2}{p} = \probchoice{t_2}{t_1}{1-p}}
    \and
    \inferrule{}{\probchoice{(\probchoice{t_1}{t_2}{p})}{t_3}{q} = \probchoice{t_1}{(\probchoice{t_2}{t_3}{\frac{(1-p)q}{1-pq}})}{pq}}
  \end{mathpar}
  The first two laws state that full probability is deterministic.
  The next, idempotence, says a coin flip between the same choice is unnecessary.
  The next, symmetry, states that probabilistic choice is symmetric under reordering of branches (with inverted probabilities).
  The last, quantitative associativity, allows for nesting coin flips in an arbitrary order (with adjusted probability). Note that the last only applies when $pq\neq 1$.
\end{example}

\begin{example}
  [Pre-convex spaces]
  The theory of pre-convex spaces has the same operations as convex spaces but drops the idempotence and symmetry laws.
\end{example}
 
 An \emph{algebra} $A : \alg \Sigma$ for a signature $\Sigma$ is a set $|A|$ 
 which provides an interpretation $\den{\op}_A : |A|^{\arity(\op)} \to |A|$ for all operations $\op : \ops$, written $\den{\op}$ when $A$ is clear.
 Any $A : \alg \Sigma$ can be used to interpret a $\Term \Sigma n$. 
 We denote an \emph{environment} $\mathsf{Env}_A(n) := |A|^n$ to be an assignment 
 of the $n$ free variables to elements of $|A|$. 
  \begin{align*}
  &\den{\_} : \Term \Sigma n \to \mathsf{Env}_A(n) \to |A|\\
  &\den{\var \;i}(\gamma) := \gamma_i\\
  &\den{\op_i(\overrightarrow{t})}(\gamma)  := \den{\op_i}(\overrightarrow{\den{t}(\gamma)})
\end{align*}
Thus, a \emph{model} for theory $\mathcal{T}$ is an algebra $A : \alg \Sigma$ 
for the signature of $\mathcal{T}$ such that all equations hold in all environments.
\[
  A \models \mathcal{T} := 
  \forall (\eq : \eqns{}), 
  (\gamma : \mathsf{Env}_A(\mathsf{ctx}_\eq)).\; 
  \den{\lhs}(\gamma)=\den{\rhs}(\gamma)
\]

Given two algebras $A,A' : \alg \Sigma$ for a fixed $\Sigma$, 
an \emph{algebra homomorphism} is a function $f : |A| \to |A'|$ that commutes with 
all the operations of $\Sigma$.
\[
      f(\den{\op}_{A}(\overrightarrow{a}))=
      \den{\op}_{A'}(\overrightarrow{f(a)})
\]
We denote $\Alg \Sigma$ as the category of $\Sigma$ algebras and algebra homomorphisms.
$\Model{\mathcal{T}}$ is the full subcategory of $\Alg \Sigma$ consisting 
of the models of $\mathcal{T}$, the algebras $A$ where $A \models \mathcal{T}$.
For any algebraic theory $\mathcal{T}$ the forgetful functor 
$U : \Model{\mathcal{T}} \to \Set$ has a left adjoint 
$F : \Set \to \Model{\mathcal{T}}$ and from this adjunction 
we have a monad $T$ on $\Set$ given by $T = U \circ F$.
Conversely, every finitary monad on $\Set$ is the free model monad
of some algebraic theory \cite{Notions-Computation-Monads-Plotkin-Power}.

\begin{example}
  Any theory $\mathcal{T}$ has a \textit{free model} for any set $X$ which is given by 
  terms $\Term \Sigma X/~\eqns{}$ quotiented by the smallest congruence generated 
  by $\eqns$. We call this the \emph{term model} of a theory $\mathcal{T}$.
  We present $\Term{\mathcal{T}} X$ as a quotient inductive recursive type below.
  Note that for a theory without equations, this type boils down to $\Term \Sigma X$.
  \begin{mathpar}
    \inferrule 
      { x : X} 
      { \var\; x : \Term{\mathcal{T}} X}
    \and
    \inferrule
      { \op : \ops \and \forall i.\;t_i : \Term{\mathcal{T}} X}
      { \op(t_1,t_2,\ldots,t_{\arity(\op)}) :\Term{\mathcal{T}} X}
    \and  
    \inferrule 
      {\eq : \eqns \and \gamma : \mathsf{Env}_{\Term{\mathcal{T}}X}(\mathsf{ctx}_\eq)}
      {\den{\mathsf{lhs}_\eq}(\gamma) = \den{\mathsf{rhs}_\eq}(\gamma)}
  \end{mathpar} 
\end{example}
$T$ being a free model monad means that for every set $X$, $T(X)$
is a model of $\mathcal T$, with an operation
$\den{\op} : T(X)^{\arity(\op)} \to T(X)$ for every operation $\op$ of
$\Sigma$ satisfying the equations $\eqns$, together with a map
$\eta : X \to T(X)$ that is universal: for every model $M$ and function
$f : X \to |M|$ there is a unique homomorphism $!_f : T(X) \to M$ with
$!_f \circ \eta = f$. The monad structure of $T$ comes from this
universal property. Its unit is $\eta$, and its multiplication
$\mu : T^2(X) \to T(X)$ is $!_{\mathrm{id}}$, the unique homomorphism
extending the identity on $T(X)$.

\begin{example}
  For set $X$, model $M$, and function $f : X \to |M|$, 
  the unique homomorphism out of the term model is defined by recursion.
  \footnote{Equality constructors elided here}
  \begin{align*}
    & !_f : \Term{\mathcal{T}}X \to M \\
    & !_f(\var \;x) = f(x) \\
    & !_f(\op(\overrightarrow{t})) = \den{\op}_M(\overrightarrow{!_f(t)})
  \end{align*}
\end{example}
\subsection{Free Model Monads as Canonical Forms}
\label{sec:free-model-monads}
While the free model always exists by this syntactic construction, it
is not a very useful model for computation. In practice, for each
theory we seek out more explicit formulations of the free models. When
the construction is particularly concrete, we can think of these as
\emph{canonical forms} of terms in the syntactic free model.

\begin{example}
  The free model of $\thy{writer}$ has carrier $T(X) = M \times X$.
  The operations are given by:
  \begin{mathpar}
      {\mathsf{tell}_e(m,x) = (m,x)}
    \and  
      {\mathsf{tell}_n(m , x) = (n \otimes m , x)}
    \and  
      {\eta(x) = (e , x)}
  \end{mathpar}
  An element $(m,x)$ can be interpreted as a canonical syntactic term $\mathsf{tell}_m(\var\; x)$.
\end{example}
\begin{example}
  The free model of $\thy{err}$ has carrier $T(X) = X + 1$,
  with $\mathsf{err} = \mathsf{inr}(\ast)$ and $\eta(x) = \mathsf{inl}(x)$.
\end{example}

\begin{example}
  The free model of $\thy{state}$ has carrier $T(X) := S \to X \times S$. The operations are given by:
  \begin{mathpar}
 
    {\mathsf{set}_s(f) = \lambda\_.\, f(s)}
    \and 
    {\mathsf{get}(f) = \lambda s.\, f(s)(s)}
    \and
    {\eta(x) = \lambda s.\, (x, s)}
  \end{mathpar}
  An element $f$ can be interpreted as a canonical syntactic term $\mathsf{get}(s \mapsto \mathsf{set}_{s'}(\var\; x))$, where $(x, s') = f(s)$
\label{ex:state-monad}
\end{example}

\begin{example}
  The free model of $\thy{wmon}$ has carrier $T(X) = \mathsf{List}(R\times X)$
  with operations
  \begin{mathpar}
    {e = []} 
    \and
    {t \otimes t' = \mathsf{append}(t,t')}
    \and
    {\mathsf{act}_r(t) = \mathsf{map}((r',x) \mapsto (r \cdot r',x))(t)}
    \and
    {\eta(x) = [(1_R,x)]}
  \end{mathpar}
  An element $[(r,x),\ldots]$ can be interpreted as a ``linear combination'' $\mathsf{act_r}(\var\; x)\otimes \cdots$.
  We use quotes because $\otimes$ is not commutative, so the order of
  the summands matters.
\end{example}

\begin{example}
  The free model of $\thy{$R$-Semi}$ $T(X) = X \to_{\mathsf{fin}} R$
  that is, a finitely supported function into the semiring, which is $0$ at all but finitely many $x$. Operations are given by
  \begin{mathpar}
    {0(x) = 0}\and
    {(t+t')(x) = t(x)+t'(x)}\and
    {\mathsf{scale}_r(t)(x) = r \cdot t(x)}\and
    {\eta(x)(x) = 1}\and
    {\eta(x)(y) = 0 \textrm{ if $x\neq y$}}
  \end{mathpar}
  An element $f$ can be interpreted as a linear combination in a
  similar manner to weighted monoids. For weighted monoids the order
  of the terms is part of the element, since it is a list. Here it is
  not, so viewing $f$ as a particular syntactic term requires choosing
  an order on the variables $X$. This can also be viewed as a quotient of the free weighted
  monoid.
\end{example}

\begin{example}
  The free model for the theory of pre-convex spaces is given by \[T(X)
  = \{l : \mathsf{List((0,1] \times X)} \mid \sum_{(p,x) \in l}{p} =
  1\}\] Note that the side condition ensures that such lists must be
    nonempty.
    \begin{mathpar}
      {\eta(x) = [(1,x)]}\and
      {\probchoice{l_1}{l_2}{1} = l_1}\and
      {\probchoice{l_1}{l_2}{0} = l_2}\and
      {\probchoice{l_1}{l_2}{p \in (0,1)} = \mathsf{append}(\mathsf{map}(-*p)(l_1),\mathsf{map}(-*(1-p))(l_2))}
    \end{mathpar}
  An element $[(p_1,x_1),(p_2,x_2),\ldots]$ can be viewed as a syntactic term $\probchoice{\eta(x_1)}{\probchoice{\eta(x_2)}{\cdots}{p_2/(1-p_1)}}{p_1}$.
\end{example}

\begin{example}
  The free model for the theory of convex spaces is given by finite probability distributions $T(X) = \{D : X \to_{\mathsf{fin}} [0,1] \mid \sum_{x\in X}D(x) = 1\}$.
  This can be viewed as a quotient of that of pre-convex spaces. Like modules of a semiring, it can be viewed as a term if we impose an ordering on the generating set $X$.
\end{example}

\subsection{Call-by-Push-Value}
Our language of study is a simply typed call-by-push-value (CBPV)
\cite{Levy-adjunction-model} calculus parameterized by a 
finitary algebraic theory $\mathcal{T}=(\Sigma,\eqns)$.
There are two versions of this calculus that we work with:
CBPV($\mathcal{T}$), for which we will provide an operational semantics
in \Cref{sec:operational-canonicity-tree}, and CBPV$^+(\mathcal{T})$,
for which we will provide a denotational semantics in
\Cref{sec:denotational-semantics-equational-canonicity}.
CBPV$^+(\mathcal{T})$ is a strict extension of CBPV($\mathcal{T}$) with 
\emph{complex values} and \emph{complex stacks}.
The purpose of this distinction is discussed at length by Levy 
\cite{Levy-adjunction-model} but it suffices to know that 
excluding complex values preserves operational canonicity while including 
complex values and stacks results in a cleaner equational theory. 
The syntax and typing rules of the full CBPV$^+(\mathcal{T}$) are shown in 
\Cref{fig:CBPV-Syntax}, where the gray forms and rules belong to
CBPV$^+(\mathcal{T})$ only. The rules are standard
\cite{Levy-adjunction-model} and are listed for reference.
\begin{mathpar}
  \inferrule
    {}
    {\Gamma \vdash V : A}
  \and  
  \inferrule 
    {}
    {\Gamma \mid \cdot \vdash M : B}
  \and
  \inferrule
    {}
    {\Gamma \mid \bullet : B \vdash S : B'}
\end{mathpar}
These are the judgments for values, computations and stacks. Here
$\Gamma$ is a context of value variables, $A$ is a value type, and $B$
and $B'$ are computation types.
In CBPV, values are data which can be bound to variables, 
passed around, and be duplicated or discarded. 
Computations represent potentially effectful programs, and a computation
of type $FA$ performs work before returning a value of type $A$.
Stacks are a generalized form of evaluation context that represent the remainder of a 
computation and a single hole into which a computation may be plugged. 
For convenience, we present the rules using a combined computation and stack 
judgment form $\Gamma \mid \Delta \vdash M : B$ containing a \emph{stoup}
\cite{Enriched-Effect-Calculus} 
$\Delta$, which is either empty, written $\cdot$,
or a single hole $\bullet : B'$. 
Rules with an unspecified stoup $\Delta$ hold for both computation 
and stack judgment forms while an empty ($\cdot$) or full ($\bullet$) 
stoup represent a computation judgment or stack judgment respectively.
The complex stacks of CBPV$^+(\mathcal{T})$ are the gray stack forms
$\lambda x : A.\, S$, $\langle S, S'\rangle$ and $\triv$ of
\Cref{fig:CBPV-Syntax}. The typing rules for $\lambda$, pairing and $\triv$
use a stoup $\Delta^+$, which may be a hole in CBPV$^+(\mathcal{T})$ and is
always empty in CBPV$(\mathcal{T})$.

We present the substitution operations, stack operations, 
and the equational theory of the language in \Cref{fig:CBPV-Ops-Eqns}.
These are standard except for the laws related to the algebraic theory.
There are four substitution operations on our syntax; value substitution $V[\gamma]$, 
computation substitution $M[\gamma]$, stack substitution $S[\gamma]$,
and stack plugging $S[M]$. 
A substitution $\gamma : \Gamma' \Rightarrow \Gamma$ assigns to each
variable $x : A$ of $\Gamma$ a value $\Gamma' \vdash \gamma(x) : A$, and
$V[\gamma]$ replaces each variable of $V$ by its value. In
\Cref{fig:CBPV-Ops-Eqns} we name each operation by its result, so
$V[\gamma]$ there stands for the map $V, \gamma \mapsto V[\gamma]$.
We additionally have notions of composing substitutions $\delta[\gamma]$,
which substitutes $\gamma$ into each value of $\delta$,
and composing stacks $S'[S]$, the stack $S'$ with $S$ in its hole.
$\beta$ and $\eta$ laws hold for all type connectives in
CBPV$^+(\mathcal{T})$, and \Cref{fig:CBPV-Ops-Eqns} shows a fragment.

Our calculus supports the construction of basic values 
with an empty type $\zero$, unit type $\mathsf{1}$, 
coproduct type $A + A'$, and product type $A \times A'$.
Eliminating a value is considered to be a computation in CBPV, 
but CBPV+ allows elimination of values into values.
A computation $M : B$ can be considered as value, 
but only after it is \emph{suspended} via a $\thunk(M) : UB$.
For computation types, we have the unit computation type $\top$, whose
introduction form is the empty tuple $\triv$,
functions $A \to B$, computation product $B \with B'$, 
and a type representing effectful values $FA$.

In addition to the typical simply typed CBPV connectives, 
our calculus includes syntax for the operations of the given theory $\mathcal{T}$.
For any operation $\op : \ops$ and $\arity(\op)$ many computations 
$\Gamma \mid \cdot \vdash M_i : B$ of type $B$ we have a computation 
$\Gamma \mid \cdot \vdash \op(\overrightarrow{M}) : B$.
As an example, for the theory of state with two states, $0$ and $1$, we have:
\begin{mathpar}
  \inferrule
    {\Gamma \mid \cdot \vdash M : B 
    \and 
    \Gamma \mid \cdot \vdash N : B}
    {\Gamma \mid \cdot \vdash \mathsf{get}(M, N) : B}
  \and  
  \inferrule
    {\Gamma \mid \cdot \vdash M : B}
    {\Gamma \mid \cdot \vdash \mathsf{set}_0(M) : B} 
    \and  
  \inferrule
    {\Gamma \mid \cdot \vdash M : B}
    {\Gamma \mid \cdot \vdash \mathsf{set}_1(M) : B}
\end{mathpar}

Of particular interest are the laws induced by the algebraic theory $\mathcal{T}$. 
Stack plugging preserves the algebraic structure,
yielding the familiar \emph{algebraicity} (or homomorphism) property of 
algebraic effects~\cite{Alg-Effects-Plotkin-Power}. 
Finally, for every context $\Gamma$ and computation type $B$, 
the set of computations $\Gamma \mid \cdot \vdash B$ 
carries the structure of a $\Sigma$-algebra and, 
moreover, satisfies the equations of $\mathcal{T}$. 
Hence, it is a model of $\mathcal{T}$. In \Cref{fig:CBPV-Ops-Eqns},
$t[\overrightarrow{M}]$ is the computation obtained from a term $t$ by
replacing its variables with $\overrightarrow{M}$, so each equation of
$\mathcal{T}$ is a law.

\begin{figure}%
    \centering
    \[
    \begin{array}{rcl}
      A,A' & := & \zero \mid \unit \mid A + A' \mid A \times A' \mid UB \\
      B,B' & := & \top \mid A \to B \mid B \with B' \mid F A \\[6pt]
      V, W & {:=} & x \mid \unitv \mid \sigma_i V \mid (V, W) \mid \thunk\,M \\
        & \mid{} & \grayhl{$\case_0(V)\{\} \mid \case_1(V)\{W\} \mid \case_+ (V)\{ x. W \mid x. W'\} \mid \case_\times (V) \{ x, y.\, W \}$} \\
      M, N & {:=} & \ret\,V \mid x \leftarrow M; N \mid \force\,V \mid \op(\overrightarrow{M}) \\
        & \mid{} & \case_0(V)\{\} \mid \case_1(V)\{M\} \mid \case_+ (V)\{ x. M \mid x. N\} \mid \case_\times (V) \{ x, y.\, M \} \\
        & \mid{} & \triv \mid \lambda x : A.\, M \mid M\,V \mid \langle M, N\rangle \mid \pi_i M \\
      S, S' & {:=} & \bullet \mid S\,V \mid \pi_i\,S \mid x \leftarrow S; N
        \mid \grayhl{$\lambda x : A.\, S \mid \langle S, S'\rangle \mid \triv$}
    \end{array}
    \]
    \begin{mathpar}
        \inferrule*[]
          { }
          {\Gamma , x:A \vdash x : A}
        \and
        \inferrule*[]
          { }
          {\Gamma \mid \bullet:B \vdash \bullet : B}
        \and
        \inferrule*[]
          {\Gamma \vdash V : \zero}
          {\Gamma \mid \cdot \vdash \case_0(V)\{\} : B}
        \and
        \colorbox{gray!20}{$
        \inferrule*[]
          {\Gamma \vdash V : \zero}
          {\Gamma \vdash \case_0(V)\{\} : A}
        $}
        \and  
        \inferrule*[]
          { }
          {\Gamma \vdash \unitv{} : \unit}
        \and
        \inferrule*[]
          {\Gamma \vdash V : \unit \and \Gamma \mid \cdot \vdash M : B}
          {\Gamma \mid \cdot \vdash \case_1(V)\{M\} : B}
        \and
        \colorbox{gray!20}{$
        \inferrule*[]
          {\Gamma \vdash V : \unit \and \Gamma \vdash W : A}
          {\Gamma \vdash \case_1(V)\{W\} : A}
        $}
        \and
           \inferrule*[]
          { \Gamma \vdash V : A  }
          { \Gamma \vdash \sigma_1 \;V  : A + A'}
        \and  
          \inferrule*[]
          { \Gamma \vdash V : A'  }
          { \Gamma \vdash \sigma_2 \;V  : A + A'}
        \and  
        \inferrule*[]
          { 
            \Gamma \vdash V : A + A' \and
            \Gamma , x : A \mid \cdot \vdash M : B \and 
            \Gamma , x : A' \mid \cdot \vdash N : B }
          { \Gamma \mid \cdot \vdash \case_+ (V)\{ x. M \mid x. N\} : B}
        \and
        \colorbox{gray!20}{$
        \inferrule*[]
          { 
            \Gamma \vdash V : A + A' \and
            \Gamma , x : A  \vdash W : A'' \and 
            \Gamma , x : A' \vdash W' : A'' }
          { \Gamma  \vdash \case_+ (V)\{ x. W \mid x. W'\} : A''}
          $}
        \and
       \inferrule*[]
          {\Gamma \vdash V : A \and \Gamma \vdash V' : A'}
          {\Gamma \vdash (V , V') : A \times A'}
        \and 
        \inferrule*[]
          {\Gamma \vdash V : A \times A' 
          \and  
          \Gamma , x : A , y : A' \mid \cdot \vdash M : B}
          {\Gamma \mid \cdot \vdash \case_\times (V) \{ x, y. \; M \} : B }
        \and 
        \colorbox{gray!20}{$
        \inferrule*[]
          {\Gamma \vdash V : A \times A' 
          \and  
          \Gamma , x : A , y : A' \vdash W : A''}
          {\Gamma \vdash \case_\times (V) \{ x, y. \; W \} : A'' }
        $}
        \and 
       \inferrule*[]
          {\Gamma \mid \cdot \vdash M : B}
          {\Gamma \vdash \thunk\;  M : U B}
        \and
        \inferrule*[]
          {\Gamma \vdash V : U B}
          {\Gamma \mid \cdot \vdash \force \:V: B}
        \and
        \inferrule*[]
          { }
          {\Gamma \mid \Delta^+ \vdash \triv{} : \top}
        \and
        \inferrule*[]
          {\Gamma , x : A \mid \Delta^+ \vdash M : B}
          {\Gamma \mid \Delta^+ \vdash \lambda x : A.\, M : A \to B}
        \and 
        \inferrule*[]
          {\Gamma \mid \Delta \vdash M : A \to B \and \Gamma \vdash V : A}
          {\Gamma \mid \Delta \vdash MV: B}
        \and 
          \inferrule*[]
          {\Gamma \mid \Delta^+ \vdash M : B 
          \and  
          \Gamma \mid \Delta^+ \vdash N : B'}
          {\Gamma \mid \Delta^+ \vdash 
          \langle M , N \rangle :  B \with B'}
        \and
        \inferrule*[]
          {\Gamma \mid \Delta \vdash M : B \with B'}
          {\Gamma \mid \Delta \vdash \pi_1(M) : B}
        \and
        \inferrule*[]
          {\Gamma \mid \Delta \vdash M : B \with B'}
          {\Gamma \mid \Delta \vdash \pi_2(M) : B'}
        \and
        \inferrule*[]
          {\Gamma \vdash V : A}
          {\Gamma \mid \cdot \vdash \ret\; V : F A}
        \and 
        \inferrule*[]
          {\Gamma \mid \Delta \vdash M : FA 
          \and 
          \Gamma, x : A \mid \cdot \vdash N : B}
          {\Gamma \mid \Delta \vdash x \leftarrow M ; N : B\vphantom{\overrightarrow{M}}}
        \and
        \inferrule*[right=\ensuremath{\op : \ops}]
          { \overrightarrow{\Gamma \mid \cdot \vdash M : B}}
          { \Gamma \mid \cdot \vdash \op(\overrightarrow{M}): B}
    \end{mathpar}
\caption{CBPV($\mathcal{T}$) \& \grayhl{CBPV$^+(\mathcal{T})$}}
\label{fig:CBPV-Syntax}
\end{figure}
 \begin{figure}
    \centering
    \renewcommand{\arraystretch}{1.3}
    \begin{tabular}{l l}
      \multicolumn{2}{l}{\textbf{Operations}} \\
      \multicolumn{2}{l}{$V[\gamma]:  \Gamma \vdash A \to (\Gamma' \Rightarrow \Gamma)  \to  \Gamma' \vdash A$}\\
      \multicolumn{2}{l}{$M[\gamma] : \Gamma \mid \cdot \vdash B \to (\Gamma' \Rightarrow \Gamma) \to \Gamma' \mid \cdot \vdash B$}\\
      \multicolumn{2}{l}{$S[\gamma] : \Gamma \mid \bullet {: B} \vdash B' \to (\Gamma' \Rightarrow \Gamma)
      \to \Gamma' \mid \bullet {: B} \vdash B'$}\\
      \multicolumn{2}{l}{$S[M] :  \Gamma \mid \bullet : B \vdash B' \to \Gamma \mid \cdot \vdash B \to \Gamma \mid \cdot  \vdash B'$}\\
      \multicolumn{2}{l}{$\delta[\gamma]: (\Gamma' \Rightarrow \Gamma'') \to (\Gamma \Rightarrow \Gamma') \to (\Gamma \Rightarrow \Gamma'')$}\\
      \multicolumn{2}{l}{$S'[S] :  \Gamma \mid \bullet : B' \vdash B'' \to \Gamma \mid \bullet : B \vdash B' \to \Gamma \mid \bullet : B \vdash B''$}\\
      \hline
      \multicolumn{2}{l}{\textbf{$\beta/\eta$ Laws (Fragment)}} \\
      $\case_+(\sigma_1 V)\{x.M\mid y.N\}=_{{\beta+}}M[V/x]$ & 
      $\case_+(V)\{x.\sigma_1x\mid y.\sigma_2y\}=_{\eta+}V$\\
      $\force(\thunk \;M) =_{\beta U} M$&
      $\thunk(\force\;V)=_{\eta U} V$\\
      $(\lambda x : A.\, M)\,V=_{\beta\to}M[V/x]$ &
      $\lambda x : A.\, (M\,x)=_{\eta\to}M$\\
      $\pi_i(\langle M_1,M_2\rangle)=_{\beta\&}M_i$ &
      $\langle \pi_1 M,\pi_2 M\rangle=_{\eta\&}M$\\
      $x \leftarrow \ret \;V ; M=_{\beta F} M[V/x]$ &
      $x \leftarrow M ; \ret\; x\; =_{\eta F}M$\\
      \hline
      \multicolumn{2}{l}{\textbf{Algebraic Laws}} \\
      $S[\op(\overrightarrow{M})] = \op(\overrightarrow{S[M]})$&
      $\lhs[\overrightarrow{M}] = \rhs[\overrightarrow{M}]$ for $\lhs = \rhs$ in $\eqns$
    \end{tabular}
  \caption{Equational Theory of CBPV$^+(\mathcal{T}$)}
  \label{fig:CBPV-Ops-Eqns}
\end{figure}

\section{Denotational Semantics and Equational Canonicity}
\label{sec:denotational-semantics-equational-canonicity}
In this section we establish equational canonicity for CBPV$^+(\mathcal{T})$. 
Our proof is organized around a \emph{CBPV$^+$ doctrine}, an abstract notion of 
model for CBPV. We first describe the syntactic and denotational models as 
instances of this doctrine, then construct a glued logical relations model. 
Initiality of the syntactic model yields the Fundamental Theorem, from which 
equational canonicity follows.

\subsection{CBPV$^+$ Doctrine}
\label{sec:cbpv+-doctrine}

This section describes the common categorical framework
from which the syntax of CBPV$^+(\mathcal{T})$, denotational semantics, 
and equational canonicity logical relation arise. 
To keep the presentation lightweight, we describe the doctrine only for
unary CBPV$^+$. By unary CBPV, we mean a restricted version 
of the CBPV calculus for which our judgment forms only have one variable.
\begin{mathpar}
  \inferrule
    {}
    {x : A \vdash V : A'}
  \and  
  \inferrule 
    {}
    {x : A \vdash M : B}
  \and
  \inferrule
    {}
    {\bullet : B \vdash S : B'}
\end{mathpar}
Consequently, the unary calculus does not admit a function type. 
Extending the construction to arbitrary contexts requires 
enriched category theory, but otherwise introduces no fundamentally new
ideas. For our set-based models the enrichment is routine, so we
do not discuss it further. A unary CBPV$^+$ doctrine consists of a functor
\[
\mathcal O :
\mathcal V^{\mathrm{op}}
\times
\mathcal C
\to
\Model{\mathcal T}.
\]
\citet{Levy-adjunction-model} models CBPV with a profunctor
$\mathcal V^{\mathrm{op}} \times \mathcal C \to \Set$, which sends $A$ and
$B$ to the set of computations from $A$ to $B$. We replace $\Set$ by
$\Model{\mathcal T}$, so these computations form a model of
$\mathcal T$.

The easiest way to understand this definition is through a concrete 
instance from which the general pattern becomes apparent.
We begin by showing that the syntax of CBPV$^+(\mathcal T)$
quotiented by its equational theory forms an instance of a unary
CBPV$^+$ doctrine which we will call the \emph{syntactic model}.
Below, values, computations and stacks are taken up to the
equations of \Cref{fig:CBPV-Ops-Eqns}.
$\mathcal{V}$ is a category where the objects are value types $A$, 
morphisms are values $A \vdash V: A'$, and composition is given by 
value substitution $V[W]$. 
$\mathcal{C}$ is a category where the objects are computation types $B$, 
morphisms are stacks $\bullet : B \vdash S : B'$, and composition is 
composition of stacks $S'[S]$. For any value type $A$ and 
computation type $B$, we have $\mathcal{O}[A , B]$ is a 
model of $\mathcal T$ with carrier given as 
computations $A \vdash M : B$ and the operations are given by the syntactic 
operations $A \vdash \op(\overrightarrow{M}) : B$. 
The contravariant action of $\mathcal{O}$ is given by substituting
values into computations, while its covariant action is given by
plugging computations into stacks. Both operations are homomorphisms. 
With functoriality specified separately in both arguments, substitution
and stack plugging commute.
\[
  (S[M])[V]
  =
  S[M[V]].
\]

To support the CBPV type connectives, a doctrine must have suitable
\emph{type structure}. By type structure, we mean the packaging of 
the introduction and elimination rules of a type along with 
$\beta$ and $\eta$ equations. From the perspective of a CBPV$^+$ doctrine, 
type structure is simply a universal property.
For example, value products are products in the category
$\mathcal V$, characterized by the natural isomorphism
\[
  \mathcal V[\_,A\times A']
  \cong
  \mathcal V[\_,A]
  \times
  \mathcal V[\_,A'].
\]
Unpacking this definition recovers the pairing and projection
rules together with their $\beta$- and $\eta$-laws. Naturality of
the isomorphism expresses that both pairing and projections commute with
substitution, yielding laws such as
\[
(V,W)[\gamma]
=
(V[\gamma],W[\gamma])
\qquad\text{and}\qquad
\pi_i(V[\gamma])
=
(\pi_iV)[\gamma].
\]
Here $\pi_i V$ is the complex value
$\case_\times(V)\{x_1, x_2.\, x_i\}$ of CBPV$^+(\mathcal{T})$.
The universal property of every connective is in
\appref{app:type-structure}{D} of this paper.

\subsection{Denotational Semantics}

We instantiate the CBPV$^+$ doctrine using the standard adjunction model of
CBPV~\cite{Levy-adjunction-model} extended to algebraic effects. 
Value types are interpreted in
$\Set$, computation types are interpreted in the category
$\Model{\mathcal T}$ of models of the algebraic theory. 
A computation may be regarded either as a function or a homomorphism
with the correspondence provided by the free-forgetful adjunction:
\[
  \Set[\den{\Gamma},U\den{B}]
  \cong
  \mathcal O[\den{\Gamma},\den{B}]
  \cong
  \Model{\mathcal T}[F\den{\Gamma},\den{B}].
\]
The free computation type is interpreted by the free model,
while function and computation product types inherit their model
structure pointwise from their codomain.

The initiality of the syntactic model means that we have a
unique structure-preserving interpretation of CBPV$^+(\mathcal{T})$ syntax into this model. 
The recursive equations characterizing this interpretation are given in
\Cref{fig:CBPV-Den-Sem}. The interpretation being a morphism of CBPV$^+$
doctrines means it preserves substitution, stack plugging, the algebraic
operations, and all type structure.
In the figure, $\den{B} \multimap \den{B'}$ is the set of
homomorphisms from $\den{B}$ to $\den{B'}$, and $\alpha_B$ is the model
structure of $\den{B}$, giving the interpretation $\alpha^{\op}_B$ of each
operation. Sequencing $x \leftarrow M; N$ is interpreted using the free
model property. The map $a \mapsto \den{N}(\gamma, a)$ extends to a
unique homomorphism $!$ out of the free model $F\den{A}$, which is then
applied to $\den{M}(\gamma)$.
For closed terms we write $\den{\_}_v : \mathsf{Val}(A) \to \den{A}$ and
$\den{\_}_c : \mathsf{Comp}(B) \to U\den{B}$ for $V \mapsto \den{V}(\star)$ and
$M \mapsto \den{M}(\star)$.

\begin{figure}
  \centering
  \[
  \begin{array}{lcl}
    \den{A} &:& \Set \\
    \den{B} &:& \Model{\mathcal T} \\
    \den{\Gamma \vdash V : A} &:& \den{\Gamma} \to \den{A}\\
    \den{\Gamma \mid \cdot \vdash M : B} &:& \den{\Gamma} \to U(\den{B})\\
    \den{\Gamma \mid \bullet : B \vdash S : B'} &:& 
      \den{\Gamma} \to \den{B} \multimap \den{B'}
  \end{array}
  \]
  \[
  \begin{array}{lcl@{\quad}lcl}
    \multicolumn{6}{c}{\textbf{Types}} \\
    \den{\zero} &=& \emptyset &
    \den{\top} &=& (\{\star\} , \alpha_\top)\\
    \den{\unit} &=& \{\star\} &
    \den{A \to B} &=& (\den{A} \to U\den{B}, \alpha_{A \to B})\\
    \den{A + A'} &=& \den{A} \uplus \den{A'} &
    \den{B \with B'} &=& (U\den{B} \times U\den{B'} , \alpha_{B \with B'})  \\
    \den{A \times A'} &=& \den{A} \times \den{A'} &
    \den{F A} &=& (UF\den{A}, \alpha_{FA}) \\
    \den{U B} &=& U\den{B} & & & \\[4pt]
    \multicolumn{6}{c}{
    \begin{aligned}
    & \alpha_{A \to B}^\op :
      (\den{A} \to U \den{B})^{\arity(\op)} \to
      \den{A} \to U \den{B}\\
    & \alpha_{A \to B}^\op(\overrightarrow{f})(a) =
      \alpha^{\op}_B(f_1(a),\ldots,f_{\arity(\op)}(a)) \\
    & \alpha_{B \with B'}^\op :
      (U\den{B} \times U\den{B'})^{\arity(\op)} \to
      U\den{B} \times U\den{B'}\\
    & \alpha_{B \with B'}^\op (\overrightarrow{p}) =
    (\alpha^{\op}_B(\pi_1(p_1), \ldots), \alpha^{\op}_{B'}(\pi_2(p_1), \ldots))
    \end{aligned}} \\[4pt]
    \multicolumn{6}{c}{\textbf{Terms}} \\
    \den{\unitv}(\gamma) &=& \star{} &
    \den{\ret\,V}(\gamma) &=& \eta(\den{V}(\gamma)) \\
    \den{\case_1(V)\{M\}}(\gamma) &=& \den{M}(\gamma) &
    \den{\lambda x.M}(\gamma)(a) &=& \den{M}(\gamma,a) \\
    \den{\sigma_1 V}(\gamma) &=& \sigma_1(\den{V}(\gamma)) &
    \den{MV}(\gamma) &=& \den{M}(\gamma)(\den{V}(\gamma)) \\
    \den{\sigma_2 V}(\gamma) &=& \sigma_2(\den{V}(\gamma)) &
    \den{\langle M , N \rangle}(\gamma) &=& (\den{M}(\gamma), \den{N}(\gamma))\\
    \den{(V,W)}(\gamma) &=& (\den{V}(\gamma),\den{W}(\gamma)) &
    \den{\pi_1(M)}(\gamma) &=& \pi_1(\den{M}(\gamma))\\
    \den{\thunk\,M}(\gamma) &=& \den{M}(\gamma) &
    \den{\pi_2(M)}(\gamma) &=& \pi_2(\den{M}(\gamma))\\
    \den{\force\,V}(\gamma) &=& \den{V}(\gamma) &
    \den{\op(\overrightarrow{M})}(\gamma)
    &=& \alpha^{\op}_{B}(\overrightarrow{\den{M}(\gamma)}) \\[4pt]
    \multicolumn{6}{c}{
    \begin{aligned}
    \den{\case_+(V)\{x.M \mid x.N\}}(\gamma)
    &=
    \begin{cases}
    \den{M}(\gamma,W) & \den{V}(\gamma)=\sigma_1(W) \\
    \den{N}(\gamma,W) & \den{V}(\gamma)=\sigma_2(W)
    \end{cases}\\
    \den{\case_\times(V)\{x,y.M\}}(\gamma)
    &=
    \begin{cases}
    \den{M}(\gamma,W,W') & \den{V}(\gamma)=(W,W')
    \end{cases}\\
    \den{x \leftarrow M;N}(\gamma)
    &= !_{\text{curry}(\den{N})(\gamma)}(\den{M}(\gamma))
    \end{aligned}}
  \end{array}
  \]
  \caption{Denotational Semantics of CBPV$^+(\mathcal T)$, a Fragment}
  \label{fig:CBPV-Den-Sem}
\end{figure}

\subsection{Equational Canonicity Logical Relation}

We now instantiate the CBPV$^+$ doctrine with a logical relations model obtained by
\emph{gluing} predicates onto the syntactic model. Rather than
interpreting value types by sets and computation types by
models of $\mathcal T$, we enrich each with a predicate.
Let $\mathsf{Val}(A)$ and $\mathsf{Comp}(B)$ be the sets of closed
values of type $A$ and closed computations of type $B$. A value predicate
$P : \mathsf{VPred}(A)$ is a map $\mathsf{Val}(A) \to \prop$. Since
closed computations carry the structure of a $\Sigma$-algebra, a
computation predicate $Q : \mathsf{CPred}(B)$ is a map
$\mathsf{Comp}(B) \to \prop$ closed under $\Sigma$, meaning that
$Q(\op(\overrightarrow{M}))$ holds whenever every $Q(M_i)$ does.
The glued value category consists of pairs $(A,P)$ where $A$ is a value
type and $P:\mathsf{VPred}(A)$. A morphism
$(A,P)\to(A',P')$ is a value
$A\vdash V:A'$ with $P(W) \Rightarrow P'(V[W])$ for all
$W : \mathsf{Val}(A)$.
Similarly, the glued computation category consists of pairs $(B,Q)$
where $B$ is a computation type and
$Q:\mathsf{CPred}(B)$. A morphism
$(B,Q)\to(B',Q')$ is a stack
$\bullet:B\vdash S:B'$ with $Q(M) \Rightarrow Q'(S[M])$ for all
$M : \mathsf{Comp}(B)$.
Finally, for every value object $(A,P)$ and computation object $(B,Q)$,
the functor consists of computations from $A$ to $B$ that preserve
the logical predicates. Concretely,
\[
\mathcal O[(A,P),(B,Q)]
:=
\left\{
A\mid\cdot\vdash M : B
\;\middle|\;
\forall W:\mathsf{Val}(A).\;
P(W)\Rightarrow Q(M[W])
\right\}.
\]
The contravariant action of $\mathcal O$ is given by value substitution,
while the covariant action is given by stack plugging. Since
$Q:\mathsf{CPred}(B)$ is closed under the operations of $\Sigma$, the
pointwise algebra structure on computations restricts to
$\mathcal O[(A,P),(B,Q)]$, yielding another instance of the CBPV$^+$
doctrine.

As with the CBPV$^+$ doctrine itself, the construction above extends
uniformly to full contexts
\cite{A-Theory-Effects-Resources-FMM}.
By initiality of the syntactic model, there is a unique
structure-preserving interpretation into the glued model. The
interpretation of each value type and computation type is 
determined entirely by the universal properties of the glued model.

We present the resulting logical predicates explicitly in
\Cref{fig:CBPV-Eq-Canon}. The relations are standard 
and $\mathcal{C}[FA]$ is the free computation predicate
\cite{Kammar-McDermott-Lifting,CBPV-Gluing-Pedro},
inductively generated by two cases. 
Whenever the value relation $\mathcal{V}[A]$ holds at $V$, 
we have that $\mathcal{C}[FA](\ret\; V)$ holds. 
Additionally, whenever the relation $\mathcal{C}[FA]$ holds at each 
operand $M_i$ of operation $\op$, we have that 
$\mathcal{C}[FA](\op(\overrightarrow{M}))$ holds.

\begin{figure}
  \[
  \begin{aligned}
\Gamma \vDash V : A 
  &:= \forall (\gamma : \Gamma).\;
      \mathcal{G}[\Gamma](\gamma)
      \implies \mathcal{V}[A](V[\gamma])\\
\Gamma \mid \cdot \vDash M : B
  &:= \forall (\gamma : \Gamma).\;
      \mathcal{G}[\Gamma](\gamma)
      \implies \mathcal{C}[B](M[\gamma])\\[0.5em]
\mathcal{G}[\emptyset]
  &:= \top\\
\mathcal{G}[\Gamma , x : A](\gamma,V)
  &:= \mathcal{G}[\Gamma](\gamma)\land \mathcal{V}[A](V)\\[0.5em]
\mathcal{V}[\zero](V)
  &:= \bot\\
\mathcal{V}[\unit](V)
  &:= \top\\
\mathcal{V}[A + A'](V)
  &:=
  \begin{cases}
    \mathcal{V}[A](W)  & \text{if } V = \sigma_1\, W\\
    \mathcal{V}[A'](W) & \text{if } V = \sigma_2\, W
  \end{cases}\\
\mathcal{V}[A \times A'](V,V')
  &:= \mathcal{V}[A](V) \land \mathcal{V}[A'](V')\\
\mathcal{V}[U B](V)
  &:= \mathcal{C}[B](\force\,V)\\[0.5em]
\mathcal{C}[\top](M)
  &:= \top\\
\mathcal{C}[A \to B](M)
  &:= \forall (V : \mathsf{Val}(A)).\;
      \mathcal{V}[A](V) \implies \mathcal{C}[B](M\,V)\\
\mathcal{C}[B \with B'](M)
  &:= \mathcal{C}[B](\pi_1(M)) \land \mathcal{C}[B'](\pi_2(M))
  \end{aligned}
  \]
  \begin{mathpar}
    \inferrule*[]
      {\mathcal{V}[A](V)}
      {\mathcal{C}[FA](\ret\,V)\vphantom{\overrightarrow{M}}}
    \and  
    \inferrule*[right=\ensuremath{\forall \; \op : \ops}]
      {\overrightarrow{\mathcal{C}[FA](M)}}
      {\mathcal{C}[FA](\op(\overrightarrow{M}))}
  \end{mathpar}
  \caption{CBPV$^+(\mathcal{T})$ Canonicity Logical Relation}
  \label{fig:CBPV-Eq-Canon}
\end{figure}

\subsection{Fundamental Theorem}
Since the glued logical relations model is an instance of the CBPV$^+$
doctrine, the initiality of the syntactic model means that we have a unique
structure-preserving interpretation of the syntax into this model, and the
fundamental theorem is a consequence of this interpretation. For completeness, 
a direct compatibility proof for the concrete logical relation is in
\appref{app:equational-fundamental-theorem}{A}.

\begin{theorem}[Fundamental theorem]
  If $\Gamma \vdash V : A$ then $\Gamma \vDash V : A$, and if
  $\Gamma \mid \cdot \vdash M : B$ then $\Gamma \mid \cdot \vDash M : B$.
  \label{thm:equational-fundamental-theorem}
\end{theorem}
From the fundamental theorem, we receive a proof of equational canonicity. 
Of particular interest are closed computations of type $FA$ which we show 
are terms of the free algebra for the signature with variables 
given by closed values of type $A$. 
To see this, we define the function reify which quotes a free algebra term 
as a closed computation.

\begin{align*}
  &\quad\mathsf{reify} : \mathsf{Term}_\Sigma (\mathsf{Val}(A)) 
  \to \mathsf{Comp}(FA)\\
  &\quad\mathsf{reify}(\var \;t) := \ret \;t\\
  &\quad\mathsf{reify}(\op(\overrightarrow{t})) := \op(\overrightarrow{\mathsf{reify}(t)})
\end{align*}

\begin{corollary}[Equational canonicity]
  \label{corollary:eq-canon}
  For every $M : \mathsf{Comp}(FA)$ there is
  $t : \mathsf{Term}_\Sigma(\mathsf{Val}(A))$ with $M = \mathsf{reify}(t)$.

  \begin{proof}
  By the Fundamental Theorem (\Cref{thm:equational-fundamental-theorem}), we have $\mathcal{C}[FA](M)$.
  We find $t$ by induction on the proof of $\mathcal{C}[FA](M)$.
  For $M = \ret\,V$, take $t = \var\,V$. For $M = \op(\overrightarrow{M})$,
  the inductive hypotheses give $t_i$ with $M_i = \mathsf{reify}(t_i)$, so
  $\op(\overrightarrow{M}) =
  \op(\overrightarrow{\mathsf{reify}(t)})
  = \mathsf{reify}(\op(\overrightarrow{t}))$. 
\end{proof}
\end{corollary}

\begin{corollary}
  If $\den{\_}_v : \mathsf{Val}(A) \to \den{A}$ is a bijection, then $\den{\_} : \mathsf{Comp}(FA) \to \Term{\mathcal{T}}{\mathsf{Val}(A)}$ is a bijection.
  \begin{proof}
      By construction $\den{\_} : \mathsf{Comp}(FA) \to
  \Term{\mathcal{T}}{\mathsf{Val}(A)}$ is a homomorphism. By the
  universal property of the free model, there is a homomorphism
  $\den{\_}^{-1} : \Term{\mathcal{T}}{\mathsf{Val}(A)} \to
  \mathsf{Comp}(FA)$ satisfying $\den{\var\;x}^{-1} =
  \ret\,\den{x}^{-1}$ and furthermore the composition $\den{\_}\circ
  \den{\_}^{-1} = \id$ so $\den{\_}^{-1}$ is injective. It suffices,
  then, to show that $\den{\_}^{-1}$ is surjective.
  First, note that $\den{\_}^{-1} \circ q = \mathsf{reify} \circ i$ where $q : \mathsf{Term}_\Sigma(\mathsf{Val}(A)) \twoheadrightarrow \mathsf{Term}_{\mathcal T}(\mathsf{Val}(A))$ is the quotient map and $i : \mathsf{Term}_\Sigma(\mathsf{Val}(A)) \cong \mathsf{Term}_\Sigma(\den{A})$ is the functorial lifting of the bijection $\den{\_}_v$.
  By Corollary~\ref{corollary:eq-canon}, $\mathsf{reify}$ is surjective, and as a bijection and quotient map, $i,q$ are also surjective. Therefore $\den{\_}^{-1}$ is surjective as well.
\end{proof}
\end{corollary}
\section{Tree Reduction}
\label{sec:operational-canonicity-tree}
This section gives a tree-based operational semantics for the smaller
calculus CBPV$(\mathcal T)$, which omits complex values, complex stacks and
the equational theory. We prove operational canonicity and strong
normalization for it (\Cref{cor:op-canon,thm:tree-strong-norm}).

\subsection{Operational Semantics of CBPV($\mathcal{T}$)}
Our tree-based operational semantics for CBPV is given by the reduction relation 
$\mtree$ in \Cref{fig:CBPV-OpSem}. It is built from $S$-steps $\mS$. A \emph{frame} $F$ is a stack of depth
exactly one, namely $\bullet\,V$, $\pi_i\,\bullet$ or $x \leftarrow \bullet;\, N$.
Every stack of CBPV$(\mathcal T)$ (\Cref{fig:CBPV-Syntax}) is a sequence of
frames, so $S$ is either $\bullet$ or $S'[F]$. There are two kinds of $S$-steps, which each take place under an arbitrary stack $S$: for each type connective, we have a $\beta$
reduction, and we have a commuting conversion that pushes a frame under an operation, as in the small-step semantics of algebraic effects and handlers
\cite{Handlers-In-Action,Pretnar-Intro-Handlers,Bauer-Pretnar-Effect-System}. By definition,
stack plugging is a congruence for $\mS$, and operations take no S-steps.
Our tree reduction does not handle reduction of operations 
$\op(\overrightarrow{M})$. 
Instead, we leave them uninterpreted 
and allow tree reductions to operate on the operands of any operation.
This is the standard operational account of algebraic effects, in
which a program computes a tree of uninterpreted operations
\cite{Plotkin-Power-Adequacy,Generic-Op-Metatheory-Alg-Effect-Johann,Adequacy-Revisited-Kavvos}.
Note that while $\mS$ is deterministic, $\mtree$ is not, as there is no specified evaluation order under an operation.
E.g., given an operation $\op(M,N)$ with 
two reducible arguments $M \mtree M'$ and $N \mtree N'$, 
we do not specify which argument to reduce first.

 \begin{figure}[H]
   \centering
    \begin{mathpar}
        F ::= \bullet\,V \mid \pi_i\,\bullet \mid x \leftarrow \bullet;\, N
    \end{mathpar}
    \begin{mathpar}
          {S[x \leftarrow \ret \;V ; N] \mS S[N[V/x]]}
        \and
          {S[F[\op(\overrightarrow{M})]] \mS S[\op(\overrightarrow{F[M]})]}
        \and
          {S[\case_+(\sigma_i V)\{x_1 .\; M_1 \mid x_2 .\; M_2\}] \mS S[M_i[V/x_i]]}
        \and
          {S[\case_1(\unitv)\{M\}] \mS S[M]}
        \and
          {S[\case_\times(V,W)\{x, y.\; M \}] \mS S[M[V/x,W/y]]}
        \and
          {S[(\lambda x : A.\, M)V] \mS S[M[V/x]]}
        \and
          {S[\pi_i \langle M_1 , M_2 \rangle] \mS S[M_i]}
        \and
          {S[\force (\thunk \; M)] \mS S[M]}
    \end{mathpar}
    \begin{mathpar}
        \inferrule*[]
          {M \mS M'}
          {M \mtree M'}
        \and
        \inferrule*[]
          {M_i \mtree M_i'}
          {\op(\ldots,M_i,\ldots) \mtree \op(\ldots,M_i',\ldots)}
    \end{mathpar}
\caption{CBPV($\mathcal{T}$) Tree Reduction $\mtree$}
\label{fig:CBPV-OpSem}
\end{figure}

\begin{lemma}[Progress and determinism]\label{lem:progress}
For every $M : \mathsf{Comp}(FA)$, $M = \ret\,V$ or $M = \op(\overrightarrow{M})$ or $\exists!\, M'.\; M \mS M'$.
\end{lemma}
\begin{proof}
By induction on $M$, inspecting the rules of \Cref{fig:CBPV-OpSem}.
\end{proof}

Note that soundness of the operational semantics with respect to the
denotational semantics follows by the same reasoning as
\Cref{sec:denotational-semantics-equational-canonicity}.
CBPV$(\mathcal T)$ is a fragment of CBPV$^+(\mathcal T)$, so its terms
are interpreted by \Cref{fig:CBPV-Den-Sem}, and each reduction is a
provable equality in the equational theory of
\Cref{fig:CBPV-Ops-Eqns}, which this interpretation validates.
\begin{theorem}[Soundness of $\mtree$]
  \label{thm:soundness-mtree}
  For $M, M' : \mathsf{Comp}(B)$, if $M \mtree M'$ then $\den{M} = \den{M'}$.
\end{theorem}

\subsection{Logical Relation}
We now adapt our logical relation from the equational to the
operational setting. As in the equational setting, for every value
type we define a predicate on closed values, and for every computation
type we define a predicate on closed computations. The only difference
is that computation predicates must now be closed not only under
the operations of $\Sigma$, but also under anti-reduction along
S-steps.
Explicitly, any $Q: \mathsf{CPred}(B)$ satisfies the following closure conditions:
\begin{mathpar}
  \inferrule*[right=\ensuremath{\forall\;\op:\ops}]
    {\overrightarrow{Q(M)}}
    {Q(\op(\overrightarrow{M}))}
  \and
  \inferrule*
    {Q(M') \and M \mS M'}
    {Q(M).\vphantom{\overrightarrow{M}}}
\end{mathpar}

We present the logical relation in \Cref{fig:CBPV-Op-Canon}. 
Its value predicates, and its computation predicates at $\top$,
$A \to B$ and $B \with B'$, are given by the clauses of
\Cref{fig:CBPV-Eq-Canon}, now on the closed terms of CBPV$(\mathcal T)$.
Compared with the equational setting, the only additional clause is that the 
free computation predicate $\mathcal{C}[FA]$ is closed under anti-reduction along
S-steps.

\begin{figure}
  \begin{mathpar}
    \inferrule*[]
      {\mathcal V[A](V)}
      {\mathcal C[FA](\ret\,V)\vphantom{\overrightarrow{M}}}
    \and
    \inferrule*[right=\ensuremath{\forall \; \op : \ops}]
      {\overrightarrow{\mathcal C[FA](M)}}
      {\mathcal C[FA](\op(\overrightarrow M))}
    \and
    \colorbox{gray!20}{$
    \inferrule*[]
      {\mathcal C[FA](M') \and M \mS M'}
      {\mathcal C[FA](M).\vphantom{\overrightarrow{M}}}
      $}
  \end{mathpar}
  \caption{CBPV$(\mathcal{T})$ Canonicity Logical Relation, the free computation predicate. The other clauses are those of \Cref{fig:CBPV-Eq-Canon}.}
  \label{fig:CBPV-Op-Canon}
\end{figure}

\begin{lemma}[Anti-reduction for computation logical predicates]
  For every computation type $B$, if $M \mS M'$ then
  $\mathcal C[B](M') \implies \mathcal C[B](M)$.
\end{lemma}
\begin{proof}
  By induction on $B$. The case $\top$ is trivial. For $A \to B$, given
  $V$ with $\mathcal V[A](V)$, we have $M\,V \mS M'\,V$ since $\mS$ is closed
  under stacks, so $\mathcal C[B](M'\,V)$ gives $\mathcal C[B](M\,V)$ by the
  inductive hypothesis. The case $B \with B'$ is the same with the stack
  $\pi_i\,\bullet$. For $FA$ it is the anti-reduction clause of
  \Cref{fig:CBPV-Op-Canon}.
\end{proof}

\begin{lemma}[Congruence for computation logical predicates]
  For every computation type, $B$, $\mathcal C[B]$ is closed under the operations of the theory.
\end{lemma}
\begin{proof}
  By induction on $B$. Trivial for $\top$ and $FA$ cases. For $A \to B$ consider $\op(\overrightarrow{M})$ where we have $\mathcal C[A \to B](M_i)$ for all $i$. Given $\mathcal V[A](V)$ we need to show $\mathcal C[B](\op(\overrightarrow{M})\,V)$. We have $\mathcal C[B](M_i\,V)$ for each $i$, so $\mathcal C[B](\op(\overrightarrow{M\,V}))$ by the inductive hypothesis. We have $\op(\overrightarrow{M})\,V \mS{} \op(\overrightarrow{M\,V})$ by the commuting conversion rule, so the result follows by anti-reduction for $B$. The $B_1\with B_2$ case is analogous.
\end{proof}

\subsection{Fundamental Theorem}
\begin{theorem}[Fundamental theorem]
  \label{thm:operational-fundamental-theorem}
  If $\Gamma \vdash V : A$ then $\Gamma \vDash V : A$, and if
  $\Gamma \mid \cdot \vdash M : B$ then $\Gamma \mid \cdot \vDash M : B$.
  \begin{proof}    
    The compatibility lemmas follow the same structure as those for
    \Cref{thm:equational-fundamental-theorem} except that usage of $\beta$
    equations and of the law $S[\op(\overrightarrow{M})] = \op(\overrightarrow{S[M]})$
    is replaced with the fact that computation predicates are
    closed under anti-reduction along S-steps. Every case is in \appref{app:fundamental-lemma}{B}. To demonstrate we consider the case for bind.
    We show that $\mathcal{C}[FA](M)$ and
    $\forall V.\, \mathcal V[A](V) \implies \mathcal{C}[B](N[V/x])$ imply
    $\mathcal{C}[B](x \leftarrow M; N)$, by induction on the proof of
    $\mathcal{C}[FA](M)$. There are three cases:
    \begin{enumerate}
    \item ($M = \ret\;V$ with $\mathcal V[A](V)$) Then $x \leftarrow \ret\;V; N \mS N[V/x]$ and follows by assumption and anti-reduction.
    \item ($M = \op(\overrightarrow{M})$ with $\overrightarrow{\mathcal C[FA](M)}$) Then $x \leftarrow \op(\overrightarrow{M}); N \mS{} \op(\overrightarrow{x\leftarrow M; N})$, and the result follows by the inductive hypothesis, the congruence lemma and anti-reduction.
    \item ($M \mS M'$ with $\mathcal C[FA](M')$) Then $(x \leftarrow M; N) \mS (x \leftarrow M'; N)$, since $\mS$ is closed under stacks, and the result follows by inductive hypothesis and anti-reduction.
    \end{enumerate}
  \end{proof}
\end{theorem}

While our logical relation concerns only $\mS$, we can establish operational canonicity of $\mtree$ as a corollary.
\begin{corollary}[Operational canonicity]\label{cor:op-canon}
  For every $M : \mathsf{Comp}(FA)$ there is
  $t : \mathsf{Term}_\Sigma(\mathsf{Val}(A))$ with $M \mtree^* \mathsf{reify}(t)$.
  \begin{proof}
  We proceed by induction on the proof of $\mathcal{C}[FA](M)$ which
  we have by \Cref{thm:operational-fundamental-theorem}.

  \begin{enumerate}
  \item $\ret V$. Then $M = \mathsf{reify}(\var\;V)$.
  \item $\op(\overrightarrow{M})$. Then by inductive hypothesis, we have $\overrightarrow{M \mtree^* \mathsf{reify}(t)}$. Since $\op$ has finite arity, we can use finitely many congruences to get $\op(\overrightarrow{M}) \mtree^* \op(\overrightarrow{\mathsf{reify}(t)}) = \mathsf{reify}(\op(\overrightarrow{t}))$.
  \item $M \mS M'$. By inductive hypothesis $M' \mtree^* \mathsf{reify}(t)$, and the result follows by transitivity, since an S-step is a tree step.
  \end{enumerate}
\end{proof}
\end{corollary}

Since the only nondeterminism in our operational semantics arises from the arbitrary order in which arguments of an operation are independently evaluated, this can be strengthened to a strong normalization result, in which every reduction sequence terminates.
\begin{corollary}[Strong normalization for $\mtree$]
  \label{thm:tree-strong-norm}
  For every $M : \mathsf{Comp}(FA)$, every reduction sequence from
  $M$ is finite.
\end{corollary}
\begin{proof}
  By induction on the proof that $\mathcal{C}[FA](M)$:
  \begin{enumerate}
  \item $\ret V$ trivial
  \item $M = \op(\overrightarrow{M})$. An operation takes no S-step, so every step from $M$ is by congruence in one of its arguments, and $M$ stays an operation. Therefore, because $\op$ has finite arity, an infinite reduction sequence would give an infinite sequence for at least one $M_i$. Since by inductive hypothesis each $M_i$ is strongly normalizing, this is a contradiction.
  \item $M \mS M'$. By \Cref{lem:progress} this is the only step from $M$, and the result follows by inductive hypothesis.
  \end{enumerate}
\end{proof}

Without equations this strengthens to reducing to the denotation itself, by operational canonicity and soundness (\Cref{thm:soundness-mtree}).
\begin{corollary}
  For any theory $\mathcal{T}$ with no equations, any type $A$ such that
  $\den{\_}_v : \mathsf{Val}(A) \cong \den{A}$ (e.g., any finite type
  $1 + 1 + \cdots {+}\, 0$), and any $M : \mathsf{Comp}(FA)$, $M \mtree^*
  \mathsf{reify}(\den{M})$.
\end{corollary}
For theories with equations, the corresponding result is the
adequacy of configuration reduction (\Cref{thm:adequacy}). Configuration
reduction computes in the free model itself.

\section{Configuration Reduction}
\label{sec:config-reduction}

The tree-based operational semantics leaves algebraic operations uninterpreted. 
In this section, we provide a second, more abstract machine-like operational 
semantics to our calculus, under the conditions that the free model of the theory is given by
a monadic container.

\subsection{Reducing Operations}
We begin with the guiding example of writer. With the writer effect,
programs are monoidally updating an accumulator variable. A natural
operational semantics then has as states a pair $\langle m, N\rangle$ where $m : M$
is an element of the monoid we are using for the accumulator and $N :
\mathsf{Comp}(FA)$ is a syntactic term that is being evaluated. To execute a term $N$, we first place it in the initial configuration, given by $\langle e , N\rangle$. Then
in addition to tree reductions that take internal steps $\langle m , N \rangle \mapsto \langle m , N' \rangle$, we
also should have a step that \emph{interprets} the syntactic operations in $N$ to update the accumulator:
\[ \langle m , \mathsf{tell}_n(N)\rangle \mapsto \langle m \otimes n , N \rangle\]
Monad aficionados will notice that the states of this machine are
precisely the carrier of the free model of the writer theory introduced in
\Cref{sec:free-model-monads} $T(\mathsf{Comp}(FA)) = M \times
\mathsf{Comp}(FA)$. The initial configuration is exactly the unit of
the monad, and the reduction rule is an instance of the multiplication of the
monad.

  This intuition carries over to our general approach.
  In operational semantics, a runtime configuration
  typically consists of some data describing the current state of the machine,
  alongside some number of syntactic subterms.
  Here, our configurations are represented by elements of the free
  model of the theory. To single out
  one subterm, we take the free model to be a \emph{monadic container}
  \cite{Uustalu-ContainerCombinatorics,Purdy-Damato-MonadicContainers},
  whose elements are a \emph{shape}, the state of the machine, together
  with a subterm at each of the shape's \emph{positions}. A step then
  either reduces a subterm or interprets an effect, updating the
  state. The effect
  rules are computed from the container, as for writer above
  (\Cref{prop:effect-reduct}), and each step
  preserves the denotation (\Cref{thm:soundness-mT}).

This establishes a sound operational semantics, but does not immediately establish strong normalization. We identify a simple sufficient condition for the operational semantics to be strongly normalizing, which we call \emph{affinity}.
Affinity says that an operational step never duplicates the subterms around the one it reduces. Assuming affinity, every reduction sequence
terminates, by a simple measure built from tree reduction (\Cref{thm:sn-T}). Affinity holds in all of our example theories whose free models are monadic containers. 

Not every free model monad is a monadic container. The free model for the theory of join-semilattices is the finite
  powerset monad, which is not polynomial. We discuss these cases further in \Cref{sec:non-container}.

\subsection{Monadic Containers}

Containers (also known as \emph{polynomial functors})
\cite{polynomial-GAMBINO_KOCK_2013,Containers-ABBOTT20053}
provide a simple combinatorial language for describing syntax and data structures. 
\begin{definition}[Container]
  A \emph{container} $\cont{\Sh}{P}$ consists of a set $\Sh$ of
  \emph{shapes} together with, for each shape $s$, a set $P(s)$ of its
  \emph{positions}. It is \emph{finitely branching} if every $P(s)$ is
  finite.
\end{definition} 

Signatures (\Cref{sec:Syntax}) are equivalent to finitely branching containers,
  though we intentionally avoid conflating them. The operations can be considered to be
  the shapes, and the arity of an operation describes a finite set of positions. This
  is a common use for containers, to describing raw syntax. However, we will use
  containers to describe the free model of an effect
  theory rather than its signature, and the two typically differ. For
  errors, the signature has one shape, $\mathsf{err}$, with no
  positions, while the free model $X + 1$ has two shapes, one with no
  positions and one with a single position.

A container determines an endofunctor on $\Set$
via the interpretation
\[ \den{\cont{\Sh}{P}}(X) := \Sigma_{s : \Sh}\, X^{P(s)} \]
We write $T$ for $\den{\cont{\Sh}{P}}$, which will be the free
model monad of \Cref{sec:free-model-monads}, and call its elements
\textit{configurations}. Each has the form
$\langle s, \overrightarrow{x}\rangle$, a shape $s$ together with a
function $\overrightarrow{x} : P(s) \to X$, which we can think of as a
vector of subterms. We call $x_p$ the
\textit{subterm} at position $p$. The functorial action of $T$ on a function $f : X \to Y$ applies
$f$ to every subterm,
\[ T(f)\langle s, \overrightarrow{x}\rangle = \langle s, f \circ \overrightarrow{x}\rangle. \]
The shape and the number of subterms are unchanged.

For configurations of computations, $X = \mathsf{Comp}(FA)$, we write
$\langle s, \overrightarrow{N}\rangle$. In typical operational semantics, a configuration should have only
finitely many subterms to evaluate, so we only use finitely branching
containers. These may still have infinitely many shapes, like writer for
infinite $M$.

The free models form a monad, so what we require is a monad whose underlying functor is a container. Such monads have a combinatorial description called a \emph{monadic container} \cite{Uustalu-ContainerCombinatorics}. Every monad
structure on $\den{\cont{\Sh}{P}}$ arises from exactly one monadic
container.
\begin{definition}[Monadic container \cite{Uustalu-ContainerCombinatorics,Purdy-Damato-MonadicContainers}]
  A \emph{monadic container} on $\cont{\Sh}{P}$ consists of
  \begin{itemize}
  \item a \emph{unit shape} $\iota : \Sh$,
  \item a \emph{flattening} operation, giving for each shape $s$ and
    family of shapes $v : P(s) \to \Sh$ the shape $\sigma(s, v)$ obtained
    by placing $v(p)$ at each position $p$ of $s$, and
  \item for each such $s$ and $v$, a \emph{position map}
    $\pr_{s,v} : P(\sigma(s,v)) \to \Sigma_{p : P(s)}\, P(v(p))$,
    written $\pr = (\pr_1, \pr_2)$, giving for each position $q$ of
    $\sigma(s, v)$ the outer position $\pr_1 q$ and the inner position
    $\pr_2 q$ it came from,
  \end{itemize}
  satisfying the unit and associativity laws
  \cite[Def.~8]{Purdy-Damato-MonadicContainers}.
  Its monad \cite[Def.~9]{Purdy-Damato-MonadicContainers} is
  $T = \den{\cont{\Sh}{P}}$ with
  \[ \eta(x) = \langle \iota, \lambda \_.\, x\rangle, \qquad
     \mu\langle s, \overrightarrow{x}\rangle
     = \langle \sigma(s, \pi_1 \circ \overrightarrow{x}), \lambda q.\, \pi_2(x_{\pr_1 q})(\pr_2 q)\rangle. \]
\end{definition}

\begin{definition}[Operational model]
  An \emph{operational model} for an algebraic theory $\mathcal T$ is
  a finitely branching monadic container $(\cont{\Sh}{P}, \iota, \sigma, \pr)$ whose monad
  $T$ is the free model monad of $\mathcal T$.
\end{definition}
\begin{example}\label{ex:writer-container}
  For writer over a monoid $(M, \otimes, e)$, the free model
  $T(X) = M \times X$ of \Cref{sec:free-model-monads} is the monadic
  container with shapes $\Sh = M$, one position $P(m) = \{\ast\}$ per
  shape, $\iota = e$, $\sigma(m, v) = m \otimes v(\ast)$ and
  $\pr(\ast) = (\ast, \ast)$
  \cite[Ex.~14]{Purdy-Damato-MonadicContainers}. A configuration
  $\langle m, N\rangle$ is an accumulator $m$ and the one term $N$ being
  run. Its unit is $\eta(N) = \langle e, N\rangle$ and
  $\mu\langle m, \langle n, N\rangle\rangle = \langle m \otimes n, N\rangle$.
\end{example}

An operational model need not be a \emph{polynomial monad} in the sense of
\citet{polynomial-GAMBINO_KOCK_2013}. Polynomial monads are exactly the
\emph{Cartesian} monadic containers, those with $P(\iota) \cong 1$
and with every $\pr : P(\sigma(s,v)) \to \Sigma_{p : P(s)} P(v(p))$ a
bijection
\cite{Uustalu-ContainerCombinatorics,Altenkirch-Pinyo-MonadicContainers}.
Writer, errors, weighted monoids and pre-convex spaces are polynomial
monads and state is the only one of our examples that is not. The unit
of the state monad has one position for each possible initial state
($P(\iota) = S$) and its multiplication drops unreached computations.

\subsection{Reduction}\label{sec:reduction}

  We define a reduction relation on configurations $\tilde t : T(\mathsf{Comp}(FA))$ whose subterms are closed CBPV computations of a returner type $F A$.
Intuitively, the shape of such a configuration specifies a state of the machine with the subterms being the terms that are currently executing in some nondeterministic order.
Matching most common operational semantics, we define a stepping relation that reduces the machine one subterm at a time.

To single out one subterm we use \emph{one-hole contexts}, given by
the derivative of the container \cite{Derivative-McBride2001,AAGM-Derivatives}:
\[ \dT(X) = \Sigma_{s : \Sh}\, \Sigma_{\pbul : P(s)}\, X^{P(s)\setminus\{\pbul\}}. \]
A context $C = \langle s, \pbul, \overrightarrow{x}\rangle$ is a shape $s$ with a hole
at position $\pbul$ and a subterm $x_p$ at every other position $p$.
Plugging $x : X$ into the hole gives the configuration
$C[x] : T(X)$ with $x$ at position $\pbul$. Like $T$, $\dT$ is a functor,
applying $f$ to the subterms around the hole, and plugging is natural:
\[ T(f)(C[x]) = \dT(f)(C)[f(x)]. \]

We define the configuration reduction relation $\mT$ in \Cref{fig:config-reduction}.
In both rules, $C : \dT(\mathsf{Comp}(FA))$ is a one-hole context.
The Pure rule simply lifts the underlying 
CBPV tree reduction $\mtree$ through an arbitrary one-hole context of the 
current configuration, allowing computations to continue reducing 
independently of the surrounding effect structure.
The Effect rule interprets a single algebraic operation. Its
left-hand side $C[S[\op(\overrightarrow{M})]]$ is a context $C$ around one subterm,
which is an operation $\op(\overrightarrow{M})$ under a stack $S$. The idea is to
interpret the syntactic operation by its semantic counterpart
$\den{\op} : T(X)^{\arity(\op)} \to T(X)$. Since the operation sits under a stack,
we first push the stack into each argument, giving
$\den{\op}(\overrightarrow{\eta(S[M])}) : T(\mathsf{Comp}(FA))$.
The remaining subterms, those of $C$, are injected into $T$ by $\eta$,
giving $\dT(\eta)(C)$. Plugging the interpreted operation into this
context yields an element of $T(T(\mathsf{Comp}(FA)))$, which $\mu$
flattens into a configuration.
\begin{figure}[!h]
  \centering
  \begin{mathpar}
    \inferrule*[right=Pure]
      {M \mtree M'}
      {C[M] \mT C[M']}
  \and 
  \inferrule*[right=Effect]
  { }
  {C[S[\op(\overrightarrow{M})]] \mT
    \mu\big(\dT(\eta)(C)\,[\den{\op}(\overrightarrow{\eta(S[M])})]\big)}
  \end{mathpar}
  \caption{Configuration Reduction $\mT$}
  \label{fig:config-reduction}
\end{figure}

Since $\mu$ is computed by $\sigma$ and $\pr$, the reduct of the
Effect rule can be computed from the monadic container operations and
the configuration each operation creates. This is how we obtain the
rules of \Cref{sec:examples}.
  In terms of monadic
containers, the reduct is most easily expressed with a derived notion of
``one-place'' flattening.
  \begin{definition}[One-place flattening]\label{def:one-place-flattening}
  Let $(\cont{\Sh}{P}, \iota, \sigma, \pr)$ be a monadic container, and
  $s : \Sh$ a shape, $\pbul : P(s)$ and $\sbul : \Sh$. The \emph{one-place flattening} of
  $\sbul$ into $s$ at $\pbul$ is
  \[ \sigmab(s, \pbul, \sbul) := \sigma(s, v)
     \qquad\text{where}\qquad
     v(p) = \begin{cases} \sbul & p = \pbul\\ \iota & p \neq \pbul \end{cases} \]
  and its \emph{one-place position map}
  $\prb : P(\sigmab(s, \pbul, \sbul)) \to P(\sbul) + (P(s) \setminus \{\pbul\})$
  says where each position of the new shape comes from, a position of
  $\sbul$ ($\inl$) or a position of $s$ other than the hole ($\inr$).
  \[ \prb(q) := \begin{cases} \inl(\pr_2 q) & \pr_1 q = \pbul\\
                              \inr(\pr_1 q) & \pr_1 q \neq \pbul \end{cases} \]
  \end{definition}

  Applying $\den{\op}$ to the argument numbers $1, \ldots, n$ gives a
  configuration
  \[ \lceil\op\rceil = \den{\op}(\eta(1),\ldots,\eta(n)) = \langle s_\op, \overrightarrow{\ell}\,\rangle : T(\{1,\ldots,n\}), \]
  with shape $s_\op$ and argument number $\ell_r$ at position $r$.
  An Effect step on $C[S[\op(\overrightarrow{M})]]$ puts $\lceil\op\rceil$ in
  the hole of $C = \langle s, \pbul, \overrightarrow{N}\rangle$.
  The new shape is
  $\sigmab(s, \pbul, s_\op)$ and its subterms are $\theta \circ \prb$, where
  $\theta : P(s_\op) + (P(s) \setminus \{\pbul\}) \to \mathsf{Comp}(FA)$
  sends $\inl\, r$ to $S[M_{\ell_r}]$ and $\inr\, p$ to $N_p$.
\[
\begin{tikzpicture}[scale=0.87, baseline=(current bounding box.center)]
\node (root) at (0,1) {$s$};
\node (a1) at (-2,-1) {$N_p$};
\node at (-.7,-.2) {$\ldots$};
\node (ai) at (0,-1) {$S[\op(\overrightarrow{M})]$};
\node at (.7,-.2) {$\ldots$};
\node (a4) at (2,-1) {$N_{p'}$};
\draw (root) -- (a1); \draw (root) -- (ai); \draw (root) -- (a4);
\end{tikzpicture}
\;\mT\;
\begin{tikzpicture}[scale=0.87, baseline=(current bounding box.center)]
\node (root) at (0,1) {$\sigmab(s, \pbul, s_\op)$};
\node (b1) at (-2,-1) {$N_p$};
\node at (-.7,-.2) {$\ldots$};
\node (bq) at (0,-1) {$S[M_{\ell_r}]$};
\node at (.7,-.2) {$\ldots$};
\node (b2) at (2,-1) {$N_{p'}$};
\draw (root) -- (b1); \draw (root) -- (bq); \draw (root) -- (b2);
\end{tikzpicture}
\]
The Pure rule is unchanged, and \Cref{fig:monadic-container-effect} states the Effect rule in container form.
\begin{figure}[!h]
  \centering
  \begin{mathpar}
  \inferrule*[right=Effect]
      {C = \langle s, \pbul, \overrightarrow{N}\rangle 
        \and \lceil\op\rceil = \config{s_\op}{\overrightarrow{\ell}}
        \\ \theta(\inl\, r) = S[M_{\ell_r}]
        \and \theta(\inr\, p) = N_p
      }
      {C[S[\op(\overrightarrow{M})]] \mT \config{\sigmab(s, \pbul, s_\op)}{\theta \circ \prb}}
  \end{mathpar}
  \caption{The Effect Rule, Stated with Monadic Container Operations}
  \label{fig:monadic-container-effect}
\end{figure}

  \begin{proposition}\label{prop:effect-reduct}
  The Effect rules of \Cref{fig:config-reduction} and
  \Cref{fig:monadic-container-effect} are equivalent. For
  $C = \langle s, \pbul, \overrightarrow{N}\rangle : \dT(\mathsf{Comp}(FA))$
  and $\theta$ as in \Cref{fig:monadic-container-effect},
  \[ \mu\big(\dT(\eta)(C)\,[\den{\op}(\overrightarrow{\eta(S[M])})]\big)
     = \config{\sigmab(s, \pbul, s_\op)}{\theta \circ \prb}. \]
  \end{proposition}
  \begin{proof}
  Since $T$ is the free model monad, $T(f)$ is a homomorphism for every
  $f$. Taking $f(i) = S[M_i]$ gives
  $\den{\op}(\overrightarrow{\eta(S[M])}) = T(f)(\lceil\op\rceil)
   = \langle s_\op, (S[M_{\ell_r}])_{r}\rangle$.
  Plugging this into $\dT(\eta)(C)$ puts $s_\op$ at the hole and
  $\eta(N_p) = \langle \iota, (N_p)\rangle$ at every other position $p$.
  Computing $\mu$ with $\sigma$ and $\pr$ then gives shape
  $\sigmab(s, \pbul, s_\op)$, and position $q$ holds the subterm at position
  $\pr_2\,q$ of the configuration placed at $\pr_1\,q$. By the case split
  defining $\prb$, this is $\theta(\prb\,q)$.
  \end{proof}

For writer (\Cref{ex:writer-container})
$\sigmab(m, \ast, n) = m \otimes n$ and
$\lceil\mathsf{tell}_n\rceil = \langle n, 1\rangle$. So the
Effect step from $\langle m, S[\mathsf{tell}_n(M)]\rangle$ gives
$\config{\sigmab(m, \ast, n)}{\theta \circ \prb}
 = \langle m \otimes n, \theta(\inl\,\ast)\rangle
 = \langle m \otimes n, S[M]\rangle$,
exactly the rule we started with.

Neither $\sigmab$ nor $\prb$ mentions terms. They use only the monadic
container, so the new shape and its position map do not depend on the
syntax of the language. Only the Pure and Effect rules refer to CBPV,
through tree reduction and stacks, so we expect configuration reduction
to carry over to call-by-value and call-by-name in the same way.

\subsection{Examples}\label{sec:examples}

When instantiated with particular monadic containers, this abstract definition reproduces familiar operational semantics, with the exception of state. \Cref{fig:example-containers} gives, for each example, the container, its monad structure and the configuration $\lceil\op\rceil$ of each operation. By \Cref{prop:effect-reduct} they determine every Effect rule, and each
example below derives its rules from one row of the table. The abstract configuration notation is sometimes clunky, and we use example-specific notation that is isomorphic to it.

\begin{figure}[h]
  \centering%
  \newcommand{\cell}[1]{\begin{tabular}[t]{@{}l@{}}#1\end{tabular}}%
  \begin{tabular}{@{}lll@{}}
    & container and monad structure & $\lceil\op\rceil$ \\ \hline
    \rule{0pt}{2.6ex}errors
      & \cell{$\Sh = \{\mathsf{run}, \mathsf{err}\}$, $P(\mathsf{run}) = \{\ast\}$, $P(\mathsf{err}) = \emptyset$\\
              $\iota = \mathsf{run}$, $\sigma(\mathsf{run}, v) = v(\ast)$, $\sigma(\mathsf{err}, v) = \mathsf{err}$\\
              $\pr(q) = (\ast, q)$}
      & \cell{$\lceil\mathsf{err}\rceil = \langle\mathsf{err}\rangle$} \\[4pt] \hline
    \rule{0pt}{2.6ex}writer
      & \cell{$\Sh = M$, $P(m) = \{\ast\}$, $\iota = e$\\
              $\sigma(m, v) = m \otimes v(\ast)$, $\pr(\ast) = (\ast, \ast)$}
      & \cell{$\lceil\mathsf{tell}_n\rceil = \langle n, 1\rangle$} \\[4pt] \hline
    \rule{0pt}{2.6ex}state
      & \cell{$\Sh = 2 \to 2$, $P(s) = 2$, $\iota = \mathrm{id}$\\
              $\sigma(s, v)(b) = v(b)(s(b))$, $\pr(b) = (b, s(b))$}
      & \cell{$\lceil\mathsf{get}\rceil = \langle \mathrm{id}, (0, 1)\rangle$\\
              $\lceil\mathsf{set}_{b'}\rceil = \langle \lambda\_.\, b', (1, 1)\rangle$} \\[4pt] \hline
    \rule{0pt}{2.6ex}weighted
      & \cell{$\Sh = \mathsf{List}(R)$, $P(l) =$ the entries of $l$, $\iota = [1]$\\
              $\sigma(l, v) = r_1 v(1) \mathbin{+\!\!+} \cdots \mathbin{+\!\!+} r_n v(n)$ for $l = [r_1, \ldots, r_n]$\\
              $\pr(q) = (i, j)$ for entry $j$ of block $i$}
      & \cell{$\lceil\mathsf{act}_r\rceil = \langle [r], 1\rangle$\\
              $\lceil e\rceil = \langle [\,]\rangle$\\
              $\lceil \otimes \rceil = \langle [1, 1], (1, 2)\rangle$} \\[4pt] \hline
    \rule{0pt}{2.6ex}pre-convex
      & \cell{lists of weights in $(0,1]$ summing to $1$\\
              structure as for weighted}
      & \cell{$\lceil {+_q} \rceil = \langle [q, 1-q], (1, 2)\rangle$\\
              \quad for $0 < q < 1$\\
              $\lceil {+_1} \rceil = \langle [1], 1\rangle$, $\lceil {+_0} \rceil = \langle [1], 2\rangle$}
  \end{tabular}
  \caption{Container data and the configuration $\lceil\op\rceil$ of each operation. Writer is worked out in \Cref{ex:writer-container} and \Cref{sec:reduction}.}
  \label{fig:example-containers}
\end{figure}

\begin{example}
  For errors, the free model is just $T(X) = X + 1$. This is the
  exception monadic container \cite[Ex.~13]{Purdy-Damato-MonadicContainers}
  \cite{Uustalu-ContainerCombinatorics}. Its shape $\mathsf{err}$
  represents the error and $\mathsf{run}$ the running computation. A
  program starts as $\langle \mathsf{run}, M\rangle$. Flattening
  $\lceil\mathsf{err}\rceil$ into its one position gives
  $\sigmab(\mathsf{run}, \ast, \mathsf{err}) = \mathsf{err}$, which has no
  positions:
  \[ \langle \mathsf{run}, S[\mathsf{err}]\rangle \mT \langle \mathsf{err}\rangle{} \]
\end{example}

\begin{example}
  For state, for concreteness we consider two states, $0$ and $1$.
  The free model $(2\times X)^2 \cong 2^2\times X^2$ is the state
  monadic container \cite[Ex.~12]{Purdy-Damato-MonadicContainers}
  \cite{Uustalu-ContainerCombinatorics}. A configuration runs the program
  for \emph{all} initial states at once. Position $b$ holds the term
  being executed when the initial state was $b$, and $s(b)$ is the
  current state of that execution. We tabulate the two, writing
  $\langle s, \overrightarrow{N}\rangle$ as
  $\langle (b_0, N_0) \mid (b_1, N_1)\rangle$ with $b_0 = s(0)$ and
  $b_1 = s(1)$. The initial configuration is
  $\eta M = \langle (0,M) \mid (1,M)\rangle{}$. At position $0$,
  flattening $\lceil\op\rceil$ updates the state $b_0$ to $s_\op(b_0)$ and picks
  the operand at position $b_0$ of $\lceil\op\rceil$. So $\mathsf{get}$ keeps
  $b_0$ and picks $S[M_{b_0}]$, and $\mathsf{set}_{b'}$ moves to $b'$ and picks
  $S[M]$. Position $1$ is symmetric:
  \begin{mathpar}
    \langle (b_0 , S[\mathsf{get}(M_0,M_1)]) \mid (b_1, N) \rangle \mT
    \langle (b_0 , S[M_{b_0}]) \mid (b_1, N) \rangle{}

    \langle (b_0 , S[\mathsf{set}_{b'}(M)]) \mid (b_1 , N) \rangle \mT
    \langle (b', S[M]) \mid (b_1 , N) \rangle{}
  \end{mathpar}
  So each pair $(b, N)$ runs on its own, and the pair at position $b$ is
  exactly the ordinary stateful machine started in state $b$.
  \Cref{sec:future-work} sketches how one Effect rule could give both
  machines.
\end{example}

\begin{example}
  The weighted monoid has free model $T(X) = \mathsf{List}(R\times X)$. As a monadic container it is a weighted version of the list monadic container \cite[Ex.~11]{Purdy-Damato-MonadicContainers}.
  If we write this configuration as $\langle (r_1,M_1) \mid{} \cdots \rangle$, then the initial configuration is $\langle (1 , M) \rangle$. Each term in $\langle (r_1,M_1) \mid \cdots \rangle$ carries a weight. When the term in a pair $(r, N)$ performs $\op$, that pair is replaced by the pairs of $\lceil\op\rceil$, with their weights multiplied by $r$. For $\mathsf{act}_{r'}$ that is one pair of weight $r \cdot r'$, for $e$ no pairs, and for $\otimes$ two pairs of weight $r$:
  \begin{mathpar}
    \langle \cdots \mid (r, S[\mathsf{act}_{r'}(M)]) \mid \cdots \rangle \mT
    \langle \cdots \mid (r\cdot r', S[M]) \mid \cdots \rangle{}\and

    \langle \cdots \mid (r, S[e]) \mid \cdots \rangle \mT
    \langle \cdots \mid \cdots \rangle{}\and

    \langle \cdots \mid (r, S[M \otimes M']) \mid \cdots \rangle \mT
    \langle \cdots \mid (r, S[M]) \mid (r, S[M']) \mid \cdots \rangle{}
  \end{mathpar}
\end{example}

\begin{example}
  The pre-convex set has a free model that can be seen as a special case of the previous where the weights are nonzero probabilities that must sum to $1$. It is the sub-container of the weighted one for the monoid $(0,1]$ under multiplication, whose shapes are the lists of weights summing to $1$. Similarly, a pair $(p, S[\probchoice{M}{N}{q}])$ is replaced by the
  pairs of $\lceil {+_q}\rceil$, with their weights multiplied by $p$. Weights
  are nonzero, so $\lceil {+_q}\rceil$ has two pairs, of weights $q$ and
  $1-q$, only when $0 < q < 1$, while $\lceil {+_1}\rceil$ keeps just $M$ and
  $\lceil {+_0}\rceil$ just $N$. That split gives the three rules:
  \begin{mathpar}
    \langle \cdots \mid (p, S[\probchoice{M}{N}{1}]) \mid \cdots \rangle \mT
    \langle \cdots \mid (p, S[M]) \mid \cdots \rangle{}

    \langle \cdots \mid (p, S[\probchoice{M}{N}{0}]) \mid \cdots \rangle \mT
    \langle \cdots \mid (p, S[N]) \mid \cdots \rangle{}
    
    \langle \cdots \mid (p, S[\probchoice{M}{N}{q}]) \mid \cdots \rangle \mT
    \langle \cdots \mid (pq, S[M]) \mid (p(1-q), S[N]) \mid \cdots \rangle
    \quad (0 < q < 1)
  \end{mathpar}
\end{example}

\paragraph{Nonexamples}\label{sec:non-container}
The free-module monad of a semiring is in general
not a monadic container. For instance, the modules of
the Boolean semiring are the join-semilattices, whose free model is
given by the finite powerset monad $\mathcal P_f(X)$. The finite
powerset functor is not a container functor, since container functors
preserve pullbacks \cite{Kock-DataTypesSymmetries} and $\mathcal P_f$
does not. Intuitively, the functorial action $T(f)$ of a container never
changes the number of subterms, but for $\mathcal P_f$, applying $f$ to
$\{x, y\}$ gives $\{f(x)\}$ when $f(x) = f(y)$. The convex spaces, whose free models
are the finitely supported probability distributions, fail for the
same reason.

The compromise is that we can give an operational model to modules of
a semiring by using the theory of weighted monoids, because the theory of
modules of a semiring is a quotient of the theory of weighted monoids. This means
that there will be multiple weighted monoid configurations that
represent the same element of the free module, e.g. $\langle (1, M) \mid
(1 , M) \rangle$ and $\langle (2 , M) \rangle$, but this can be viewed
as an equivalence that can be imposed on the final result
configurations that arise from the operational execution. A more
systematic treatment of such quotients is a possible direction for
future work (\Cref{sec:future-work}).

\subsection{Nondeterminism and Soundness}\label{sec:soundness-config}

Note that configuration reduction $\mT$ is nondeterministic in several ways. First, we have the parallel nondeterminism of $\mtree$ where subterms of an operation can be evaluated in any order. We have the same kind of nondeterminism with the Pure and Effect rules, whose context $C$ can single out any subterm of the configuration, so the subterms of a configuration can be evaluated in any order. As our weighted monoid example shows, this kind of nondeterminism can be desirable, as the weighted monoid encodes a form of nondeterminism and so it makes sense that the evaluation between different nondeterministic choices has no specified order of evaluation. Similarly for our state example, there is no need to evaluate the term for different start states in a particular order. There is also a nondeterminism between Pure and Effect steps. When a subterm is an operation under a stack, $S[\op(\overrightarrow{M})]$, we might either take Pure steps or an Effect step. The Effect rule interprets the operation at once, while Pure steps first move it out of $S$ one frame at a time.

We show that all of this nondeterminism is ultimately benign. Every
step preserves the denotational semantics (\Cref{thm:soundness-mT}). For
an affine operational model, every reduction sequence terminates
(\Cref{thm:sn-T}), and at ground types every run reaches the same final
configuration (\Cref{thm:adequacy}).

Because we derived the operational semantics and denotational semantics from the free model, we can show that they agree by an abstract categorical calculation.
First, we define a denotation of a configuration as the Kleisli extension of the 
denotation of closed computations of type $FA$:
\begin{align*}
  &\den{\_}_T : T(\mathsf{Comp}(FA)) \to T(\den{A})\\
  &\den{\tilde t}_T = \mu(T(\den{\_}_c)(\tilde t))
\end{align*}

Then the denotation of the initial configuration is correct.
\begin{theorem}[Soundness of initial configuration]
  $\den{\eta(M)}_T = \den{M}_c$
\end{theorem}
\begin{proof}
  Immediate by the unit and naturality laws of the monad
  $$\mu(T(\den{\_}_c)(\eta(M))) = \mu(\eta(\den{M}_c)) = \den{M}_c$$
\end{proof}

And the denotation is an invariant of the reduction relation:
\begin{theorem}[Soundness of $\mT$]
  For $\tilde{t}, \tilde{t'} : T(\mathsf{Comp}(FA))$, if $\tilde{t} \mT \tilde{t'}$ then $\den{\tilde{t}}_T = \den{\tilde{t'}}_T$.
  \begin{proof}
    \begin{enumerate}
    \item The Pure case follows from the soundness of tree reduction (\Cref{thm:soundness-mtree}).
    \item The Effect case. Both sides of the Effect rule denote
      $\mu\big(\dT(\den{\_}_c)(C)\,[\den{\op}(\overrightarrow{\den{S[M]}_c})]\big)$.
      This follows from the monad laws, since stacks and $\den{\_}_T$ are
      homomorphisms. The calculation is in \appref{app:soundness-config}{C}.
    \end{enumerate}
  \end{proof}
  \label{thm:soundness-mT}
\end{theorem}

\subsection{Termination and Adequacy}\label{sec:termination}

To prove that configuration reduction terminates, we need one more
condition. Every subterm of the reduct of an Effect step is either an
operand $S[M_i]$ or a neighboring subterm of the configuration. Operands
may be copied, but neighboring subterms must not be. We call this
condition
\emph{affinity}\footnote{This is unrelated to Jacobs' \emph{affine
monads} \cite{Jacobs-AffineMonads}, those with $T(1) \cong 1$. Writer,
errors and state are affine in our sense, yet $T(1)$ is $M$, $1 + 1$
and $S \to S$ respectively.}.
\begin{definition}[Affinity]
  A monadic container is \emph{affine} if plugging a shape into a
  hole never copies a neighbor. Formally, for all $s, \sbul : \Sh$ and
  $\pbul : P(s)$,
  \[ \pr_1 : P(\sigmab(s, \pbul, \sbul)) \to P(s) \text{ is injective on } \{\, q \mid \pr_1\, q \neq \pbul \,\}. \]
  An operational model is affine if its monadic container is.
\end{definition}

\begin{example}
  Writer, errors and state are affine, since $\pr_1$ is injective
  everywhere. For writer and errors every shape has at most one position.
  For state, $\pr(b) = (b, s(b))$, so $\pr_1$ is the identity. The
  $\mathsf{get}$ steps of state only drop the branch not taken and copy
  nothing.
\end{example}
\begin{example}
  The weighted monoid and the pre-convex sets are affine, since each
  neighbor receives $\iota = [1]$, which has a single position. Only the
  hole can split, as in the $\otimes$ step.
\end{example}

To prove that an affine operational model ensures termination, we show that every step decreases a \emph{termination measure} in a well-founded order. A computation is measured by how many tree steps and operations it still has to perform (using \Cref{thm:tree-strong-norm}). For a configuration, we collect the measures of its subterms, one for each position, into a multiset.

\begin{definition}[Termination measure]
  \label{def:metric}
  Let $M : \mathsf{Comp}(FA)$, define a measure $|M|$ as follows:
  \begin{enumerate}
  \item If $M = \ret\; V$, then ${|\ret\; V|} = 0$.
  \item If $M = \mathsf{op}(\overrightarrow{M})$ then $|\mathsf{op}(\overrightarrow{M})| = 1 + {\sum_{i} |M_i|}$.
  \item Otherwise $M$ takes exactly one S-step $M \mS N$
    (\Cref{lem:progress}), and $|M| = 1 + |N|$.
  \end{enumerate}
  Because $\mtree$ is strongly normalizing (\Cref{thm:tree-strong-norm}), this measure is finite.

  For a configuration $\tilde t = \langle s, \overrightarrow{N}\rangle : T(\mathsf{Comp}(FA))$,
  we define the multiset $|\tilde t| = \{\, |N_p| \mid p : P(s) \,\}$.
  Multisets are compared by the Dershowitz--Manna ordering, where
  $Y <_M X$ if $Y = (X - R) + A$ for some nonempty $R \subseteq X$ such that every element of $A$
  is smaller than some element of $R$. Because $<$ on $\mathbb N$ is well founded, so is $<_M$
  \cite{Dershowitz-Manna}.
\end{definition}

Then we get strong normalization because every step reduces the measure.
\begin{theorem}[Strong normalization for $\mT$]\label{thm:sn-T}
  Let the operational model be affine and $\tilde{t} : T(\mathsf{Comp}(FA))$.
  (1) Every reduction sequence from $\tilde{t}$ is finite.
  (2) If no step applies to $\tilde{t}$, then $\tilde{t} = T(\ret)(\tilde{v})$
  for some $\tilde{v} : T(\mathsf{Val}(A))$.
  \begin{proof}
    Every reduction $\tilde{t} \mT \tilde{t}'$ decreases the measure of \Cref{def:metric}, $|\tilde{t}'| <_M |\tilde{t}|$.

    If every element of $|\tilde{t}|$ is $0$, then every subterm has measure $0$. Since an
    operation or a term that steps has measure at least $1$, every subterm
    is of the form $\ret\; V$ (\Cref{lem:progress}), so $\tilde{t} = T(\ret)(\tilde{v})$.

    Otherwise, some element of $|\tilde{t}|$ is positive. By \Cref{lem:progress}
    the corresponding subterm is an operation or takes an S-step, so at least
    one reduction applies.
    Consider any $\tilde{t} \mT \tilde{t}'$. There are two cases:
    \begin{enumerate}
    \item If it is a Pure step, it rewrites one subterm $N_p \mtree N'_p$, and
      $|N'_p| < |N_p|$ by induction on the tree step, using case 3 of
      \Cref{def:metric} for an S-step and case 2 for a step under
      an operation. So $R = \{|N_p|\}$ and $A = \{|N'_p|\}$. The others stay the same.
    \item If it is an Effect step from $C[S[\op(\overrightarrow{M})]]$ with
      $C = \langle s, \pbul, \overrightarrow{N}\rangle$, then by \Cref{prop:effect-reduct} each
      subterm of the reduct is a neighbor $N_p$ ($p \neq \pbul$) or an
      operand $S[M_i]$, and by affinity each neighbor occurs at most once.
      So we take $R = \{|S[\op(\overrightarrow{M})]|\}$ plus the dropped neighbors, and
      $A$ the measures of the operands in the reduct. It suffices that
      $|S[M_i]| < |S[\op(\overrightarrow{M})]|$ for each $i$. If
      $S$ has $k$ frames, then $S[\op(\overrightarrow{M})]$ reaches
      $\op(\overrightarrow{S[M]})$ in $k$ S-steps, one per frame. So by
      \Cref{lem:progress} and cases 3 and 2 of \Cref{def:metric},
      $|S[\op(\overrightarrow{M})]| = k + 1 + \sum_i |S[M_i]|$.
    \end{enumerate}
    Since $<_M$ is well founded, every reduction sequence terminates.
  \end{proof}
\end{theorem}

Combining soundness and strong normalization gives computational
adequacy in the sense of Plotkin and Power \cite{Plotkin-Power-Adequacy}.
Running the configuration machine computes the denotation.
\begin{theorem}[Computational adequacy]\label{thm:adequacy}
  Let the operational model be affine, and let $A$ be a type such that
  $\den{\_}_v : \mathsf{Val}(A) \to \den{A}$ is an isomorphism (for
  instance, a ground type built from $0, 1, +, \times$). Then for every
  $M : \mathsf{Comp}(FA)$:
  (1) $\eta(M) \mT^* T(\ret)(\tilde v)$ for some $\tilde v : T(\mathsf{Val}(A))$;
  (2) $T(\den{\_}_v)(\tilde v) = \den{M}_c$; and
  (3) $\tilde v$ is unique.
\end{theorem}
\begin{proof}
  (1) follows from \Cref{thm:sn-T} with $\tilde t = \eta(M)$. For (2), by
  soundness $\den{M}_c = \den{\eta(M)}_T = \den{T(\ret)(\tilde v)}_T$, and
  \begin{align*}
    \den{T(\ret)(\tilde v)}_T
    &= \mu(T(\den{\_}_c)(T(\ret)(\tilde v)))
     = \mu(T(\eta \circ \den{\_}_v)(\tilde v))
     = T(\den{\_}_v)(\tilde v).
  \end{align*}
  (3) follows from (2). Container functors preserve injections, so
  $T(\den{\_}_v)$ is injective, and any two results with the same
  denotation $\den{M}_c$ are equal.
\end{proof}
\begin{corollary}
  Under the hypotheses of \Cref{thm:adequacy}, if $\den{M}_c = \den{N}_c$
  then $\eta(M)$ and $\eta(N)$ reduce to the same final configuration.
\end{corollary}

\section{Discussion}
We have presented a uniform framework in which both denotational and
operational semantics are derived from the same algebraic description of
effects. Starting from an algebraic theory, we construct a denotational
semantics, a tree-based operational semantics, and a configuration-based
operational semantics, together with generic canonicity, soundness and
adequacy arguments connecting them. The configuration-based semantics
is computed from the monadic container that is the free model. A short
calculation shows that every step preserves the denotation. Under a simple
affinity condition on the container, every run terminates, and at ground
types it ends in the denotation of the program. We believe this provides a foundation for developing operational
semantics for algebraic effects in a systematic
manner.

\subsection{Related Work}

\subsubsection{Operational Semantics of Algebraic Effects}
Many operational accounts of algebraic effects
\cite{Plotkin-Power-Adequacy,Generic-Op-Metatheory-Alg-Effect-Johann,Adequacy-Revisited-Kavvos} proceed by evaluating programs 
to effect trees (or free monadic syntax), which are subsequently interpreted 
into a semantic model via a monad morphism or handler. This cleanly separates 
operational evaluation from denotational interpretation, but does not,
in general, account for the equations of an algebraic theory during reduction.
By contrast, our configuration reduction contains tree reduction as its
Pure steps (\Cref{fig:config-reduction}), so a program can be evaluated to a
tree and then interpreted, or have its operations interpreted as they are
reached. Either way, configuration reduction is sound with respect to the
denotational semantics (\Cref{thm:soundness-mT}).

The work on Guarded Interaction Trees
\cite{Guarded-Interaction-Trees} shares our idea
that operational reduction should take place within semantic
configurations. 
Programs are interpreted in the semantic domain of guarded interaction trees, 
and a corresponding notion of configuration-based reduction is defined at the 
level of interaction trees. 
The authors define a small-step, configuration-based operational semantics 
for a recursive language with natural numbers and I/O, and prove it sound and 
adequate with respect to the guarded interaction tree reduction.
Their proofs use effect-specific logical relations, while ours are
effect generic.

\subsubsection{Monadic Rewriting and Operational Semantics}\label{sec:monadic-opsem}
Monadic rewriting \cite{Gavazzo-Faggian-MonadicRewriting} uses a rewriting relation between elements of a monad, and later work builds on it for studying a monadic variant of intersection types \cite{GTV-MonadicIntersection} and operational semantics for type-and-effect systems \cite{DGZ-MonadicTypeEffect}. 
Monadic rewriting is based on a rewriting relation between elements of $T(\mathsf{Exp})$, which is quite similar in spirit to our configuration reduction. Furthermore, their reductions evaluate the terms in the configuration by evaluating the operation into the monad. Our approach differs firstly in that we emphasize a notion of \emph{one-place reduction} capturing the common style of existing operational semantics, whereas in all of these works, the effects are evaluated as a \emph{parallel} reduction. That is, when an effectful step is taken, all subterms take a step simultaneously, whereas in our system a subterm is non-deterministically chosen to step. Furthermore, \citet{Gavazzo-Faggian-MonadicRewriting} and \citet{GTV-MonadicIntersection} restrict to weakly Cartesian monads, so they exclude some common examples such as reader and state. On the other hand, we require monadic containers and not all weakly Cartesian monads are monadic containers, e.g., finite multisets (\Cref{sec:future-work}).
\citet{DGZ-MonadicTypeEffect} do not restrict to weakly Cartesian monads, but
do restrict to the underlying operational semantics of terms being
deterministic, which we do not, as our tree reduction includes
non-deterministic evaluation order.
Because none of these works use one-place reduction, none use monadic
containers and none of them identify anything like an affinity
condition or prove an analogous termination result, as they include
non-terminating systems. \citet{GTV-MonadicIntersection} characterize
termination by intersection typability, but this is not analogous to
our termination result, which follows for simply typed CBPV terms.
Additionally none of these approaches prove soundness and adequacy
with respect to denotational semantics or relate to the tree-based
reduction.

\subsubsection{Structural Operational Semantics}
Our work is related in spirit to the bialgebraic approach to
structural operational semantics of Turi and Plotkin
\cite{GSOS-Turi-Plotkin}, also known as abstract GSOS, which seeks a systematic categorical account of
operational semantics. There, operational behavior is specified by a
distributive law between syntax and behavior functors, inducing a canonical
coalgebra on the initial algebra. The bialgebraic framework has been extended 
to higher-order languages \cite{GSOS-Binders}, including call-by-push-value
\cite{GSOS-CBPV}, and stateful computation \cite{GSOS-State}. 

Such a distributive law gives each syntactic constructor its behavior
from the behaviors of its arguments, so behavior sits under the syntax. We
instead fix the syntax of CBPV and derive the configurations and effect
rules from an algebraic theory and its free model. Our configurations are
the other way around, with the syntax inside the free model of the
effects, and our container is this free model, rather than a polynomial functor describing the
syntax. Closest in this line is the extension of Turi and
Plotkin's framework with algebraic effects by Abou-Saleh and Pattinson
\cite{Abou-Saleh-Pattinson}, where the one-step behavior of a program
lies in an effect monad, such as the free model of a Lawvere theory,
possibly combined with a comodel for state. Their languages are first-order
and their states hold a single program rather than many subterms. Whether
our configuration reduction can be given in their framework is open.

\subsection{Future Work}\label{sec:future-work}
  To support a wider variety of effects, we would like
  to extend this framework to include theories of local state
  \cite{StatonLocalState,StatonInstances}, whose models are algebras in
  categories of presheaves rather than in $\Set$. There computational contexts are
  indexed by varying notions of locality and resource. This perspective is
  well developed in work of Fiore on discrete generalized polynomial
  functors~\cite{Generalized-Poly-Fiore} and of Fiore and Hur on
  equational systems~\cite{Equational-Systems-Fiore-Hur},
  where presheaf categories model syntax with scope and dependency.

  Our strong normalization proof for configuration reduction relies on a 
  termination measure and therefore does not extend directly to 
  general recursion. Future work could integrate
  guarded recursion~\cite{A-Modality-Recursion-Nakano} into our framework
  by treating it as an algebraic effect.

  Our operations have finite arity, and our containers have finitely
  many positions. Allowing infinite arities, for example state over an
  infinite set of states, would give configurations with infinitely many
  subterms. The initial configuration then holds one copy of the program
  for each initial state, and unless the program is already a value, no
  finite reduction sequence reaches a normal form. Plotkin's adequacy result for infinitary algebraic
  effects \cite{Plotkin-Infinitary-Adequacy} is a natural starting
  point.

  Our configuration-based reduction is currently restricted to algebraic effects 
  whose free models are monadic containers, excluding 
  quotient-based theories such as finite sets and convex spaces. 
  A natural extension is to reduce in a theory with fewer equations
  whose free model $T'$ is a monadic container, such as lists for finite
  sets, with a split surjection $q : T'(X) \twoheadrightarrow T(X)$, allowing reduction to operate on canonical
  representatives while remaining sound with respect to the quotient equations. 
  The present framework is the special case of the trivial quotient. For 
  example, finite sets admit canonical representatives as sorted, 
  duplicate-free lists over totally ordered sets. We leave a systematic 
  development of quotient-based operational models to future work.

  Our configurations for state are not those of the ordinary
  stateful machine, which pair a single state with a single term. That
  machine instead comes from the state \emph{comodel}
  \cite{Plotkin-Power-Tensors,Uustalu-StatefulRunners}, which answers
  each $\mathsf{get}$ and $\mathsf{set}$, and more generally from
  runners \cite{Ahman-Bauer-Runners}. Both kinds of configuration are
  right modules over the monad $T$. A comodel $C$ gives the module
  $C \times -$, and our configurations are $T$ acting on itself by $\mu$.
  We believe the Effect rule extends to any right $T$-module whose
  functor is a container, with the module action in place of $\mu$,
  which would recover the ordinary machine as well as ours.

  Algebraic theories can be combined by sum and tensor
  \cite{HYLAND200670}, and we would like operational models to combine
  in the same way. Some combinations stay within monadic containers. Our
  weighted monoids are the tensor of monoids with writer, and tensoring
  with writer or with finite state preserves monadic containers in
  general. Others do not, since the tensor of monoids with itself has
  finite multisets as its free model. Distributive laws of monadic
  containers \cite{Purdy-Damato-MonadicContainers} describe when two
  monadic containers compose, and so suggest when an operational model
  can be built from its parts. Whether distributive laws of monadic
  containers preserve affinity is open.

\begin{acks}
This material is based upon work supported by the National Science Foundation CISE Graduate Fellowships (CSGrad4US) under Grant No. 2313998. Any opinions, findings, and conclusions or recommendations expressed in this material are those of the author(s) and do not necessarily reflect the views of the National Science Foundation.
\end{acks}

\section*{Disclosure of Use of AI in Research}
We used an LLM-based coding assistant to help formalize the paper's
results in an unpublished Agda mechanization. The formalization revealed a gap in one proof and
several smaller errors, which we corrected by hand.

\section*{Data Availability Statement}
This paper has no artifact, all novel theorems are proven in the paper or
\ifextended its appendices\else the appendices of its extended version\fi.

\bibliographystyle{ACM-Reference-Format}
\bibliography{popl27}

\ifextended
\appendix
\section{Proof of the Fundamental Theorem of Section~\ref{\xrp sec:denotational-semantics-equational-canonicity}}\label{app:equational-fundamental-theorem}
We prove \Cref{\xrp thm:equational-fundamental-theorem} by induction on the
typing derivation, with one case for each rule of CBPV$^+(\mathcal T)$ in
\Cref{\xrp fig:CBPV-Syntax} for the judgments $\Gamma \vdash V : A$ and
$\Gamma \mid \cdot \vdash M : B$. In each case we fix $\gamma$ with
$\mathcal G[\Gamma](\gamma)$ and apply the inductive hypotheses at $\gamma$.
Values and computations are taken up to the equations of
\Cref{\xrp fig:CBPV-Ops-Eqns}, so $\mathcal V[A]$ and $\mathcal C[B]$ hold of a
term exactly when they hold of every term equal to it. In each $\beta$ case
below, the $\beta$ law turns the goal into a hypothesis.
\begin{description}
\item[Variable.] $\mathcal V[A](\gamma(x))$ is a conjunct of
  $\mathcal G[\Gamma](\gamma)$.
\item[Unit.] $\mathcal V[\unit](\unitv)$ holds by definition.
\item[Injection.] $\mathcal V[A_1 + A_2](\sigma_i(V[\gamma]))$ is
  $\mathcal V[A_i](V[\gamma])$, the hypothesis.
\item[Pair.] $\mathcal V[A \times A'](V[\gamma], V'[\gamma])$ is the
  conjunction of the two hypotheses.
\item[Thunk.] We need $\mathcal C[B](\force(\thunk\,M[\gamma]))$. Since
  $\force(\thunk\,M[\gamma]) = M[\gamma]$, this is the hypothesis.
\item[Force.] $\mathcal V[UB](V[\gamma])$ is by definition
  $\mathcal C[B](\force\,V[\gamma])$.
\item[Case on the empty type.] The hypothesis on $V$ gives
  $\mathcal V[\zero](V[\gamma])$, which is $\bot$, so the case is vacuous.
  The same holds for the complex value $\case_0(V)\{\}$.
\item[Case on unit.] The $\eta$ law for $\unit$ gives $V[\gamma] = \unitv$.
  Then $\case_1(\unitv)\{M[\gamma]\} = M[\gamma]$, so the goal is the
  hypothesis on $M$. The complex value $\case_1(V)\{W\}$ is the same with
  $\mathcal V[A]$ in place of $\mathcal C[B]$.
\item[Case on sums.] The hypothesis on $V$ gives $V[\gamma] = \sigma_i\,W$
  with $\mathcal V[A_i](W)$. Then
  \[\case_+(\sigma_i W)\{x_1.\,M_1[\gamma] \mid x_2.\,M_2[\gamma]\} =
  M_i[\gamma][W/x_i],\]
  and $\mathcal G[\Gamma, x_i : A_i](\gamma, W)$ gives
  $\mathcal C[B](M_i[\gamma][W/x_i])$. The complex value
  $\case_+(V)\{x_1.\,W_1 \mid x_2.\,W_2\}$ is the same with $\mathcal V[A]$
  in place of $\mathcal C[B]$.
\item[Case on products.] The hypothesis on $V$ gives $W$ and $W'$ with
  $V[\gamma] = (W, W')$, $\mathcal V[A](W)$ and $\mathcal V[A'](W')$. Then
  \[\case_\times(W, W')\{x, y.\,M[\gamma]\} = M[\gamma][W/x, W'/y],\]
  and
  $\mathcal G[\Gamma, x : A, y : A'](\gamma, W, W')$ gives $\mathcal C[B]$
  of it. The complex value $\case_\times(V)\{x, y.\,W\}$ is the same with
  $\mathcal V[A]$ in place of $\mathcal C[B]$.
\item[Return.] The first rule of $\mathcal C[FA]$ in
  \Cref{\xrp fig:CBPV-Eq-Canon}.
\item[Bind.] Let $S := x \leftarrow \bullet;\, N[\gamma]$. The hypothesis on
  $M$ gives $\mathcal C[FA](M[\gamma])$, and we show $\mathcal C[B](S[M'])$
  for every $M'$ with $\mathcal C[FA](M')$, by induction on the proof of
  $\mathcal C[FA](M')$.
  \begin{itemize}
  \item ($M' = \ret\,V$ with $\mathcal V[A](V)$) Then
    $S[\ret\,V] = N[\gamma][V/x]$, and $\mathcal G[\Gamma, x : A](\gamma, V)$
    gives $\mathcal C[B](N[\gamma][V/x])$.
  \item ($M' = \op(\overrightarrow{M})$ with $\mathcal C[FA](M_i)$ for each
    $i$) The inductive hypothesis gives $\mathcal C[B](S[M_i])$ for each $i$,
    and $\mathcal C[B]$ is closed under the operations of $\Sigma$, so
    $\mathcal C[B](\op(\overrightarrow{S[M]}))$. The algebraic law
    $S[\op(\overrightarrow{M})] = \op(\overrightarrow{S[M]})$ makes this the
    goal.
  \end{itemize}
\item[Operation.] Each $\mathcal C[B](M_i[\gamma])$ holds, and
  $\mathcal C[B]$ is closed under the operations of $\Sigma$, so
  $\mathcal C[B](\op(\overrightarrow{M[\gamma]}))$.
\item[Empty tuple.] $\mathcal C[\top](\triv)$ holds, since $\mathcal C[\top]$
  holds of every computation.
\item[Abstraction.] Given $\mathcal V[A](V)$, we have
  $(\lambda x : A.\, M[\gamma])\,V = M[\gamma][V/x]$, and
  $\mathcal G[\Gamma, x : A](\gamma, V)$ gives
  $\mathcal C[B](M[\gamma][V/x])$.
\item[Application.] The hypothesis on $M$, applied to $V[\gamma]$, which
  satisfies $\mathcal V[A]$ by the hypothesis on $V$.
\item[Pairing.] By definition of $\mathcal C[B_1 \with B_2]$, we need
  $\mathcal C[B_i](\pi_i \langle M_1[\gamma], M_2[\gamma]\rangle)$ for each $i$. Since
  \[\pi_i \langle M_1[\gamma], M_2[\gamma]\rangle = M_i[\gamma],\] this is the hypothesis on $M_i$.
\item[Projection.] $\mathcal C[B_i](\pi_i\,M[\gamma])$ is a conjunct of
  $\mathcal C[B_1 \with B_2](M[\gamma])$.
\end{description}

\section{Proof of the Fundamental Theorem of Section~\ref{\xrp sec:operational-canonicity-tree}}\label{app:fundamental-lemma}
We prove \Cref{\xrp thm:operational-fundamental-theorem} by induction on the
typing derivation, with one case for each rule of CBPV$(\mathcal T)$ in
\Cref{\xrp fig:CBPV-Syntax}. In each case we fix $\gamma$ with
$\mathcal G[\Gamma](\gamma)$ and apply the inductive hypotheses at $\gamma$.
Each $\beta$ rule of \Cref{\xrp fig:CBPV-OpSem} with $S = \bullet$ is an S-step,
so in the $\beta$ cases the anti-reduction lemma for computation logical
predicates turns the goal into a hypothesis.
\begin{description}
\item[Variable.] $\mathcal V[A](\gamma(x))$ is a conjunct of
  $\mathcal G[\Gamma](\gamma)$.
\item[Unit.] $\mathcal V[\unit](\unitv)$ holds by definition.
\item[Injection.] $\mathcal V[A_1 + A_2](\sigma_i(V[\gamma]))$ is
  $\mathcal V[A_i](V[\gamma])$, the hypothesis.
\item[Pair.] $\mathcal V[A \times A'](V[\gamma], V'[\gamma])$ is the
  conjunction of the two hypotheses.
\item[Thunk.] We need $\mathcal C[B](\force(\thunk\,M[\gamma]))$. Anti-reduction
  along $\force(\thunk\,M[\gamma]) \mS M[\gamma]$ gives it from the hypothesis.
\item[Force.] $\mathcal V[UB](V[\gamma])$ is by definition
  $\mathcal C[B](\force\,V[\gamma])$.
\item[Case on the empty type.] The hypothesis on $V$ gives
  $\mathcal V[\zero](V[\gamma])$, which is $\bot$, so the case is vacuous.
\item[Case on unit.] The only closed value of type $\unit$ is $\unitv$, so
  $V[\gamma] = \unitv$. Then $\case_1(\unitv)\{M[\gamma]\} \mS M[\gamma]$,
  and anti-reduction gives $\mathcal C[B]$ of it from the hypothesis on $M$.
\item[Case on sums.] The hypothesis on $V$ gives $V[\gamma] = \sigma_i\,W$
  with $\mathcal V[A_i](W)$. Then
  \[\case_+(\sigma_i W)\{x_1.\,M_1[\gamma] \mid x_2.\,M_2[\gamma]\} \mS
  M_i[\gamma][W/x_i],\]
  and $\mathcal G[\Gamma, x_i : A_i](\gamma, W)$ gives
  $\mathcal C[B](M_i[\gamma][W/x_i])$. Anti-reduction finishes the case.
\item[Case on products.] The hypothesis on $V$ gives $W$ and $W'$ with
  $V[\gamma] = (W, W')$, $\mathcal V[A](W)$ and $\mathcal V[A'](W')$. Then
  \[\case_\times(W, W')\{x, y.\,M[\gamma]\} \mS M[\gamma][W/x, W'/y],\]
  and
  $\mathcal G[\Gamma, x : A, y : A'](\gamma, W, W')$ gives $\mathcal C[B]$
  of the reduct. Anti-reduction finishes the case.
\item[Return.] The first rule of $\mathcal C[FA]$ in
  \Cref{\xrp fig:CBPV-Op-Canon}.
\item[Bind.] Let $S := x \leftarrow \bullet;\, N[\gamma]$. The hypothesis on
  $M$ gives $\mathcal C[FA](M[\gamma])$, and we show $\mathcal C[B](S[M'])$
  for every $M'$ with $\mathcal C[FA](M')$, by induction on the proof of
  $\mathcal C[FA](M')$.
  \begin{itemize}
  \item ($M' = \ret\,V$ with $\mathcal V[A](V)$) Then
    $S[\ret\,V] \mS N[\gamma][V/x]$, and $\mathcal G[\Gamma, x : A](\gamma, V)$
    gives $\mathcal C[B](N[\gamma][V/x])$. Anti-reduction finishes the case.
  \item ($M' = \op(\overrightarrow{M})$ with $\mathcal C[FA](M_i)$ for each
    $i$) The frame rule gives
    $S[\op(\overrightarrow{M})] \mS \op(\overrightarrow{S[M]})$. The
    inductive hypothesis gives $\mathcal C[B](S[M_i])$ for each $i$, so the
    congruence lemma for computation logical predicates gives
    $\mathcal C[B](\op(\overrightarrow{S[M]}))$. Anti-reduction finishes the
    case.
  \item ($M' \mS M''$ with $\mathcal C[FA](M'')$) Since $\mS$ is closed under
    stacks, $S[M'] \mS S[M'']$. The inductive hypothesis gives
    $\mathcal C[B](S[M''])$, and anti-reduction finishes the case.
  \end{itemize}
\item[Operation.] Each $\mathcal C[B](M_i[\gamma])$ holds, so the congruence
  lemma for computation logical predicates gives
  $\mathcal C[B](\op(\overrightarrow{M[\gamma]}))$.
\item[Empty tuple.] $\mathcal C[\top](\triv)$ holds, since $\mathcal C[\top]$
  holds of every computation.
\item[Abstraction.] Given $\mathcal V[A](V)$, we have
  $(\lambda x : A.\, M[\gamma])\,V \mS M[\gamma][V/x]$, and
  $\mathcal G[\Gamma, x : A](\gamma, V)$ gives
  $\mathcal C[B](M[\gamma][V/x])$. Anti-reduction finishes the case.
\item[Application.] The hypothesis on $M$, applied to $V[\gamma]$, which
  satisfies $\mathcal V[A]$ by the hypothesis on $V$.
\item[Pairing.] By definition of $\mathcal C[B_1 \with B_2]$, we need
  $\mathcal C[B_i](\pi_i \langle M_1[\gamma], M_2[\gamma]\rangle)$ for each $i$. Since
  \[\pi_i \langle M_1[\gamma], M_2[\gamma]\rangle \mS M_i[\gamma],\] anti-reduction gives it from the hypothesis on $M_i$.
\item[Projection.] $\mathcal C[B_i](\pi_i\,M[\gamma])$ is a conjunct of
  $\mathcal C[B_1 \with B_2](M[\gamma])$.
\end{description}

\section{Proof of Soundness of Configuration Reduction}\label{app:soundness-config}
We prove \Cref{\xrp thm:soundness-mT} by cases on the rule.
\begin{description}
\item[Pure.] This follows from the soundness of tree reduction
  (\Cref{\xrp thm:soundness-mtree}).
\item[Effect.] By the Effect rule the redex is
  $C[S[\op(\overrightarrow{M})]]$ for some $C : \dT(\mathsf{Comp}(FA))$, and the
  reduct is $\mu\big(\dT(\eta)(C)[\den{\op}(\overrightarrow{\eta(S[M])})]\big)$.
  Write $f^\dagger = \mu\circ T(f)$ for the Kleisli extension of
  $f : X \to T(Y)$, so that $\den{\_}_T = \den{\_}_c^\dagger$.
  We use four facts.
  \begin{enumerate}
  \item[(A)] $f^\dagger \circ \eta = f$ (naturality of $\eta$ and the unit law).
  \item[(B)] $f^\dagger \circ \mu = \mu \circ T(f^\dagger)$ (naturality
    of $\mu$ and associativity).
  \item[(C)] Plugging is natural, $T(f)(C[x]) = \dT(f)(C)[f(x)]$
    (\Cref{\xrp sec:reduction}).
  \item[(D)] $f^\dagger$ is a homomorphism, being the unique
    homomorphism extending $f$ (\Cref{\xrp sec:algebraic-theories}), so it
    commutes with $\den{\op}$.
  \end{enumerate}
  For the left-hand side, (C) gives
  \[ \den{C[S[\op(\overrightarrow{M})]]}_T
     = \mu\big(T(\den{\_}_c)(C[S[\op(\overrightarrow{M})]])\big)
     = \mu\big(\dT(\den{\_}_c)(C)\,[\den{S[\op(\overrightarrow{M})]}_c]\big). \]
  For the right-hand side,
  \begin{align*}
    &\den{\mu(\dT(\eta)(C)[\den{\op}(\overrightarrow{\eta(S[M])})])}_T\\
    &\quad= \mu\big(T(\den{\_}_c^\dagger)(\dT(\eta)(C)[\den{\op}(\overrightarrow{\eta(S[M])})])\big) && \text{(B)}\\
    &\quad= \mu\big(\dT(\den{\_}_c^\dagger\circ\eta)(C)\,[\den{\_}_c^\dagger(\den{\op}(\overrightarrow{\eta(S[M])}))]\big) && \text{(C)}\\
    &\quad= \mu\big(\dT(\den{\_}_c)(C)\,[\den{\op}(\overrightarrow{\den{\_}_c^\dagger(\eta(S[M]))})]\big) && \text{(A), (D)}\\
    &\quad= \mu\big(\dT(\den{\_}_c)(C)\,[\den{\op}(\overrightarrow{\den{S[M]}_c})]\big) && \text{(A)}
  \end{align*}
  The two sides differ only in the plugged value, so it suffices that
  \[ \den{S[\op(\overrightarrow{M})]}_c = \den{\op}(\overrightarrow{\den{S[M]}_c}). \]
  This holds because stacks are interpreted as homomorphisms
  (\Cref{\xrp fig:CBPV-Den-Sem}), so
  \[ \den{S[\op(\overrightarrow{M})]}_c = \den{S}(\den{\op}(\overrightarrow{\den{M}}))
   = \den{\op}(\overrightarrow{\den{S}(\den{M})})
   = \den{\op}(\overrightarrow{\den{S[M]}_c}). \]
\end{description}

\section{Type Structure of the CBPV$^+$ Doctrine}\label{app:type-structure}
\Cref{\xrp sec:cbpv+-doctrine} gives type structure by universal properties
and works out value products. \Cref{tab:type-structure} lists the
universal property of every connective of the unary calculus. The
function type needs contexts with more than one variable and is omitted,
as in \Cref{\xrp sec:cbpv+-doctrine}. The two isomorphisms for $+$ are
compatible with substitution,
$M[\case_+(z)\{x.\,W_1 \mid y.\,W_2\}/w] = \case_+(z)\{x.\,M[W_1/w] \mid y.\,M[W_2/w]\}$,
and the two for $\with$ with plugging,
$\langle S_1, S_2\rangle[M] = \langle S_1[M], S_2[M]\rangle$.
\begin{table}[h]
  \centering
  $\begin{array}{@{}l@{\qquad}l@{\qquad}l@{}}
  \text{Connective} & \multicolumn{2}{l}{\text{Universal property (natural isomorphisms)}}\\ \hline
  \rule{0pt}{2.6ex}
  \mathsf 1,\ \times & \mathcal V[\_, \mathsf 1] \cong 1 & \mathcal V[\_, A_1 \times A_2] \cong \mathcal V[\_, A_1] \times \mathcal V[\_, A_2] \\[3pt]
  \mathsf 0,\ + & \mathcal V[\mathsf 0, \_] \cong 1 & \mathcal V[A_1 + A_2, \_] \cong \mathcal V[A_1, \_] \times \mathcal V[A_2, \_] \\
   & \mathcal O[\mathsf 0, \_] \cong 1 & \mathcal O[A_1 + A_2, \_] \cong \mathcal O[A_1, \_] \times \mathcal O[A_2, \_] \\[3pt]
  U & \mathcal V[\_, UB] \cong \mathcal O[\_, B] & \\[3pt]
  F & \mathcal C[FA, \_] \cong \mathcal O[A, \_] & \\[3pt]
  \top,\ \with & \mathcal O[\_, \top] \cong 1 & \mathcal O[\_, B_1 \with B_2] \cong \mathcal O[\_, B_1] \times \mathcal O[\_, B_2] \\
   & \mathcal C[\_, \top] \cong 1 & \mathcal C[\_, B_1 \with B_2] \cong \mathcal C[\_, B_1] \times \mathcal C[\_, B_2]
  \end{array}$
  \caption{Universal properties of the connectives in a unary CBPV$^+$ doctrine.}
  \label{tab:type-structure}
\end{table}
 \fi

\end{document}